\RequirePackage{tikz}
\RequirePackage{fix-cm}
\documentclass[review]{elsarticle}

\usepackage{blindtext}
\usepackage[hidelinks]{hyperref}
\usepackage[absolute,overlay]{textpos}
\usepackage[margin=1.44in]{geometry}
\usepackage{hypernat}

\usepackage{amssymb}
\usepackage{amsmath}
\usepackage{bm}
\usepackage{shellesc}
\usepackage{subcaption}
\usetikzlibrary{arrows,shapes,decorations.pathreplacing,calligraphy}
\usepackage{booktabs}
\usepackage{adjustbox}
\usepackage{siunitx}
\usepackage{upgreek}
\usepackage[utf8]{inputenc}
\usepackage[T1]{fontenc}

\usepackage{etoolbox}
\usepackage{algorithm}
\usepackage{algpseudocode}
\usepackage{physics}
\usepackage{booktabs}
\usepackage{multirow}
\usepackage{array}
\usepackage{color}
\usepackage{colortbl}
\usepackage{longtable}
\usepackage{amsthm}

\theoremstyle{plain}

\theoremstyle{remark}

\usepackage{tikz}
\usetikzlibrary{shapes.geometric,arrows.meta,positioning,calc,fit,backgrounds}

\usepackage{pgfplots}
\pgfplotsset{compat=1.17}

\usepackage[]{caption}[2008/08/24]
\newcolumntype{L}[1]{>{\raggedright\arraybackslash}p{#1}}
\newcolumntype{C}[1]{>{\centering\arraybackslash}p{#1}}

\definecolor{accentorange}{RGB}{204,102,0}
\definecolor{tealfill}  {RGB}{225,245,238}
\definecolor{tealbd}    {RGB}{15,110,86}
\definecolor{phasebg}   {RGB}{248,248,248}
\definecolor{phaseborder}{RGB}{180,178,170}

\tikzset{
box/.style={rectangle, rounded corners=3pt, draw=black!60,
	fill=white, line width=0.7pt,
	text width=54mm, minimum height=9mm,
	font=\small, align=center},
newbox/.style={box, fill=orange!18, draw=accentorange, line width=1pt},
decision/.style={diamond, draw=black!60, fill=blue!6, line width=0.7pt,
	aspect=2.4, font=\small, align=center, inner sep=1pt},
arr/.style={-Stealth, line width=0.8pt, black!60},
newarr/.style={-Stealth, line width=0.8pt, accentorange},
label/.style={font=\footnotesize, midway},
phaselabel/.style={font=\small\itshape, text=phaseborder},
}

\journal{Journal of the Mechanics and Physics of Solids}
\begin{document}

\begin{frontmatter}
	
\title{\fontsize{11.5}{16}\selectfont
	From Low- to High-Cycle Fatigue in Short Fiber-Reinforced Polymers: A Multimechanism Phase-Field Model Driven by Viscoelastic--Viscoplastic Damage
	}

\author[1,2]{Behrouz Arash\corref{cor1}}
\ead{behrouza@oslomet.no}
\author[1]{Shadab Zakavati}
\ead{shgha8895@oslomet.no}
\author[3]{Quan Wang}
\ead{quanwang@polyu.edu.hk}
\author[4]{Timon Rabczuk}
\ead{timon.rabczuk@uni-weimar.de}

\address[1]{Department of Mechanical, Electrical and Chemical Engineering,
	Oslo Metropolitan University, Pilestredet 35, 0166 Oslo, Norway}
\address[2]{Green Energy Lab, Department of Mechanical, Electrical and
	Chemical Engineering, OsloMet -- Oslo Metropolitan University, Oslo, Norway}
\address[3]{Department of Mechanical Engineering, The Hong Kong Polytechnic University, Hung Hom, Kowloon, Hong Kong, China}	
\address[4]{Institute of Structural Mechanics, Bauhaus-Universit{\"a}t Weimar,
	Marienstra{\ss}e 15, 99423 Weimar, Germany}

\cortext[cor1]{Corresponding author}

\begin{abstract}

Estimating the fatigue life of short fiber-reinforced polymer composites in both low- and high-cycle regimes requires a description of physically different damage mechanisms affected by loading history, temperature, and microstructure. Current phase-field fatigue models account for them only in part. In this study, an anisotropic phase-field model is developed at finite deformation. Two matrix mechanisms drive the fatigue damage. Above yield, the accumulated viscoplastic strain controls low-cycle fatigue, and the accumulated non-equilibrium strain energy of the viscoelastic network controls high-cycle fatigue. A weighted sum of the two degrades the fracture toughness without requiring an explicit transition between fatigue regimes. An adaptive cycle-jump scheme extrapolates the fatigue variables across skipped cycles and limits the toughness loss in each jump. The model is validated with experimental results for a short carbon fiber-reinforced epoxy. A variance decomposition reveals that the toughness decay scale dominates the fatigue life at high stress. At low stress, it shares this role equally with the viscoelastic weight. The cycle-jump scheme predicts the explicit life within 1.4\% at low cycles and cuts the number of simulated cycles by over two orders of magnitude at high cycles. Simulations of a notched specimen indicate that fatigue life drops with loading amplitude and fiber content, rises with temperature, and is sensitive to the fiber configuration.

\end{abstract}	

	
\end{frontmatter}

\section{Introduction}
\label{sec:intro}


Short carbon fiber-reinforced polymer (SCFRP) composites have found extensive applications in aerospace, automotive, and renewable energy systems due to their high specific stiffness and strength, as well as their tunable anisotropic properties~\cite{holbery2006natural,soutis2005carbon,mortazavian2015fatigue,arash2019viscoelastic2}. In practical applications, these structures are seldom subjected to monotonic loading. Instead, they typically experience repeated mechanical loading, making fatigue a major factor governing their service life~\cite{wicaksono2023review,kovavcevic2024unifying}. Their fatigue performance is particularly sensitive to fiber orientation, fiber volume fraction, and operating temperature. Experimental characterization of these dependencies is costly and time-consuming, as each combination of parameters requires numerous specimens and loading cycles. Consequently, developing predictive models that account for these effects and capture both crack initiation and propagation is essential for reliable and cost-effective structural design.


The mechanical behavior of SCFRP composites is inherently anisotropic due to the distribution of fiber orientations. This anisotropy influences their elastic response, inelastic deformation, and fracture behavior~\cite{hente2016coarse,gusev2017finite,wang2019numerical,arash2019viscoelastic2}. The polymer matrix also exhibits rate-dependent behavior governed by interacting viscoelastic and viscoplastic mechanisms~\cite{poulain2014finite,arash2019viscoelastic,arash2021finite,arash2025phase}. These mechanisms play different roles in fatigue damage under cyclic loading. When the local stress exceeds a yield-type threshold, irreversible viscoplastic deformation accumulates and primarily drives low-cycle fatigue (LCF) at high stress amplitudes. At lower amplitudes, even below the yield threshold, the viscoelastic network experiences repeated stretching followed by incomplete recovery. The resulting accumulation of non-equilibrium strain energy becomes the dominant source of damage in high-cycle fatigue (HCF). Consequently, a purely viscoplastic formulation cannot capture fatigue degradation in this regime~\cite{schijve2003fatigue,wicaksono2023review}. A unified constitutive-fracture framework must therefore account for both fatigue regimes through their underlying viscoelastic-viscoplastic (VE-VP) mechanisms, rather than relying solely on phenomenological descriptions based on loading history.


Conventional fracture mechanics simulations have commonly employed cohesive zone models~\cite{ortiz1999finite,li2005use,zhang2022fatigue}, element deletion methods~\cite{lapczyk2007progressive}, and extended finite element methods~\cite{pike2015xfem,kastner2016xfem}. However, these techniques face challenges when describing complex crack patterns in fiber-reinforced polymers. Phase-field modeling offers an alternative variational approach in which fracture is represented by a continuous damage field~\cite{francfort1998revisiting,bourdin2000numerical,miehe2010phase}. This formulation avoids explicit crack tracking and enables mesh-independent solutions~\cite{ambati2015review,wu2020phase}. Its applications have expanded to several fracture mechanisms, including brittle~\cite{miehe2010thermodynamically,borden2012phase}, ductile~\cite{miehe2016phase,dean2026hybrid,zhen2024phase,zhang2026phase}, and cohesive fracture~\cite{verhoosel2013phase}. Further developments have incorporated anisotropic fracture energy~\cite{li2015phase,teichtmeister2017phase,clayton2015phase}, making the framework particularly suitable for fiber-reinforced composites~\cite{bleyer2018phase,gultekin2016phase,dean2020multi,lv2026anisotropic}. Recent advances have also addressed the coupled VE-VP response of polymer matrices under finite deformation. In particular, Arash and coworkers~\cite{arash2021finite,arash2023effect,arash2025phase} developed phase-field fracture formulations for finite-deformation VE-VP materials with hygrothermal coupling, covering epoxy nanocomposites and short fiber-reinforced polymers~\cite{arash2026phase}.


Phase-field approaches have also been developed to describe fatigue fracture. Most existing formulations introduce a scalar history variable that evolves under cyclic loading and reduces either the fracture toughness or the crack driving force~\cite{schreiber2020phase,grossman2022phase}. Alessi et al.~\cite{alessi2018phenomenological} and Carrara et al.~\cite{carrara2020framework} developed variational phase-field formulations for fatigue in brittle materials. Subsequent studies extended these approaches to ductile fatigue~\cite{lo2019phase,seiler2020efficient}, with several recent reviews providing comprehensive overviews of these developments~\cite{li2023review,kalina2023overview}. Further extensions have addressed fatigue in fiber-reinforced composite laminates~\cite{zhang2024fatigue,hcf2025matrix}. Of particular relevance to the present work, Dean et al.~\cite{dean2026experimental} recently introduced a phase-field framework for ductile fatigue in SCFRPs. In their formulation, fatigue degradation is represented through a phenomenological reduction in fracture toughness based on the accumulated loading history. However, this approach does not distinguish between the viscoplastic and viscoelastic contributions to fatigue damage described earlier. As a result, it cannot reproduce the transition from LCF to HCF through the material's intrinsic rate-dependent constitutive behavior. Moreover, neither this formulation nor the other fatigue models discussed above investigates how fiber content and multiple fiber-family configurations influence fatigue performance.


Computational expense presents another major challenge in fatigue simulations. Explicit cycle-by-cycle analyses become computationally prohibitive when dealing with the $10^4$ to $10^6$ cycles typically encountered in engineering applications, particularly for coupled phase-field problems involving finite deformation. To reduce this cost, cycle-jump and other acceleration techniques have been developed. These methods approximate the evolution of fatigue history variables across multiple skipped cycles and periodically solve the mechanical fields explicitly. Their effectiveness has been demonstrated for brittle and elastomeric phase-field fatigue models~\cite{loew2020accelerating,yang2023acceleration,jaccon2023adaptive,ding2024adaptive}. However, such acceleration strategies have not yet been integrated into finite-deformation VE-VP constitutive frameworks. This integration is particularly challenging because fatigue variables evolve through distinct physical mechanisms, while fracture toughness degradation must be accurately controlled throughout each cycle jump.

Building on these considerations, this study introduces an anisotropic phase-field framework for fatigue fracture in SCFRP composites under finite deformation. An adaptive cycle-jump strategy is also developed to enable computationally efficient simulations at engineering fatigue-life scales. The main contributions are as follows. First, a fatigue driving force is formulated by combining accumulated viscoplastic strain with accumulated non-equilibrium strain energy. Both contributions reduce the fracture toughness but represent distinct viscoplastic and viscoelastic fatigue mechanisms. This formulation allows their individual roles in fatigue damage evolution to be investigated, while their relative importance varies between the LCF and HCF regimes. Second, an adaptive cycle-jump algorithm is introduced for finite-deformation phase-field fatigue simulations. The method first performs an explicit block of cycles, with its length determined by a convergence criterion based on the per-cycle damage increment. The evolution rates of the two fatigue variables are then estimated and extrapolated over the skipped cycles. The resulting increments are adjusted using a scaling factor that accounts for the decay of the peak force with increasing cycle count. An additional limiter controls the fracture toughness reduction during each jump, with its formulation derived analytically in closed form. Third, the proposed model is calibrated and validated against experimental data from monotonic and fatigue tests. A full factorial study combined with variance decomposition is conducted to quantify the individual effects of fatigue parameters and their interactions on fatigue life. The accuracy and computational efficiency of the cycle-jump algorithm are further assessed through a convergence study. Fourth, the framework incorporates temperature-dependent material behavior and anisotropic fracture resistance associated with multiple fiber families at arbitrary orientations. Its predictive capabilities are demonstrated using a single-edge notched specimen. The simulations investigate the effects of loading amplitude, temperature, fiber content, and fiber configuration on fatigue crack initiation and propagation.


The remainder of this paper is structured as follows. Section~\ref{sec:model} describes the VE-VP constitutive model and the anisotropic phase-field fracture formulation. Their detailed derivations are provided in Appendices~\ref{app:fiber} to \ref{app:fem}. Section~\ref{sec:fatigue} introduces the fatigue extension and the adaptive cycle-jump algorithm, while Appendix~\ref{app:limiter} presents the derivation of the constraint on toughness reduction within each jump. Section~\ref{sec:results} reports the model calibration and validation, sensitivity analysis, convergence assessment of the cycle-jump method, and simulations of a single-edge notched specimen. Finally, Section~\ref{sec:summary} summarizes the main findings and discusses possible directions for future research.

\section{Constitutive and phase-field fracture model}
\label{sec:model}


The VE-VP constitutive model and the anisotropic phase-field formulation adopted here were developed and verified in our earlier work~\cite{arash2026phase,ARASH2027114194}. This section summarizes only the equations needed for the fatigue extension in Section~\ref{sec:fatigue}. The remaining derivations are moved to the appendices. These cover the fiber strain energy and stress (\ref{app:fiber}), the time integration of the internal variables (\ref{app:integration}), and the linearization and consistent tangent of the finite element problem (\ref{app:fem}). The new elements are the stress-based crack driving force of Eq.~\eqref{eq:energy_density} and its coupling to the fatigue-degraded toughness and crack strength introduced in Section~\ref{sec:fatigue}. Temperature is treated as a prescribed input that is uniform in space and constant in time. This assumption suits tests performed in a controlled thermal environment, and the heat conduction equation is therefore not solved.

\subsection{Viscoelastic-viscoplastic constitutive model}
\label{sec:const}

\begin{figure}[H]
	\centering
	\begin{subfigure}[b]{0.49\textwidth}
		\centering
		\includegraphics{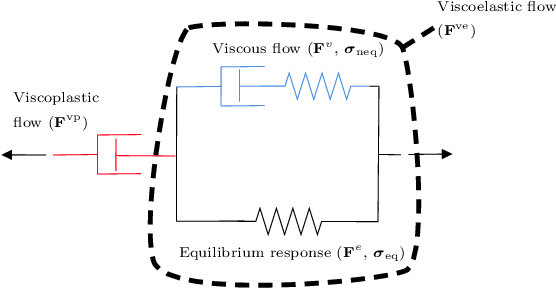}
		\caption{}
		\label{fig:rheological_model_a}
	\end{subfigure}%
	\hfill
	\begin{subfigure}[b]{0.49\textwidth}
		\centering
		\includegraphics{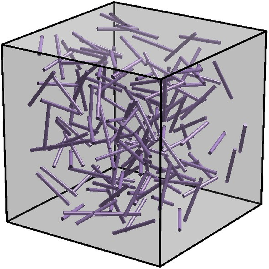}
		\caption{}
		\label{fig:rheological_model_b}
	\end{subfigure}
	\caption{(a) One-dimensional schematic of the VE-VP
		constitutive model for SCFRP composites. (b) Representative volume element
		showing the carbon fiber network embedded in the epoxy matrix.}
	\label{fig:rheological_model}
\end{figure}

The formulation combines the rate-dependent inelastic behavior of the epoxy matrix with the anisotropic reinforcement provided by multiple carbon fiber families (Fig.~\ref{fig:rheological_model}). At each material point, the state is described by three quantities. These are the deformation gradient $\mathbf{F}$, the viscous internal variable $\bar{\mathbf{F}}^v$, and the viscoplastic internal variable $\bar{\mathbf{F}}^{\text{vp}}$.

The deformation gradient is split into a volumetric and an isochoric part, and the Jacobian into a mechanical and a thermal part,
\begin{equation}
	\mathbf{F} = J^{1/3}\bar{\mathbf{F}},
	\qquad
	J = J_m J_\theta,
	\qquad
	J_\theta = 1 + \alpha_\theta (\theta - \theta_0),
	\label{eq:F_decomp}
\end{equation}
where $\det[\bar{\mathbf{F}}]=1$, $\alpha_\theta$ is the thermal expansion
coefficient, $\theta$ is the absolute temperature, and $\theta_0=296$~K is the
reference temperature. Since the temperature is uniform, $J_\theta$ is a fixed
scalar in each simulation. The isochoric part follows the multiplicative split
of~\cite{govindjee1997presentation},
\begin{equation}
	\bar{\mathbf{F}} = \bar{\mathbf{F}}^{\text{ve}}\bar{\mathbf{F}}^{\text{vp}},
	\qquad
	\bar{\mathbf{F}}^{\text{ve}} = \bar{\mathbf{F}}^e\bar{\mathbf{F}}^v,
	\label{eq:Fbar_decomp}
\end{equation}
with the associated left Cauchy-Green tensors
$\bar{\mathbf{B}}=\bar{\mathbf{F}}\bar{\mathbf{F}}^T$,
$\bar{\mathbf{B}}^{\text{ve}}=\bar{\mathbf{F}}^{\text{ve}}(\bar{\mathbf{F}}^{\text{ve}})^T$,
and $\bar{\mathbf{B}}^e=\bar{\mathbf{F}}^e(\bar{\mathbf{F}}^e)^T$.

The viscous stretching tensor follows the Argon flow rule~\cite{boyce1988large},
\begin{equation}
	\mathbf{D}^v =
	\frac{\dot{\varepsilon}^v}{\tau_{\text{neq}}}
	\operatorname{dev}\!\left[\boldsymbol{\sigma}'_{\text{neq}}\right],
	\qquad
	\dot{\varepsilon}^v = \dot{\varepsilon}_0
	\exp\!\left[\frac{\Delta H}{k_b \theta}
	\left(\!\left(\frac{\tau_{\text{neq}}}{\tau_0}\right)^{\!m} - 1\right)\right],
	\label{eq:eps_dot_v}
\end{equation}
where $\tau_{\text{neq}}=\|\operatorname{dev}[\boldsymbol{\sigma}_{\text{neq}}]\|_F$,
$\boldsymbol{\sigma}'_{\text{neq}}=\mathbf{R}_e^T\boldsymbol{\sigma}_{\text{neq}}\mathbf{R}_e$
is the non-equilibrium Cauchy stress rotated by the elastic rotation
$\mathbf{R}_e$ of the polar decomposition
$\bar{\mathbf{F}}^e=\mathbf{R}_e\mathbf{U}_e$, $k_b$ is the Boltzmann constant,
$\dot{\varepsilon}_0$ is the pre-exponential strain rate, $\Delta H$ is the
activation energy, $\tau_0$ is the athermal shear resistance, and $m$ is a
material exponent. The viscoplastic
stretching tensor is driven by the total deviatoric stress and follows a
threshold-based rule,
\begin{equation}
	\mathbf{D}^{\text{vp}}
	= \frac{\dot{\varepsilon}^{\text{vp}}}{\tau_{\text{p}}}
	\operatorname{dev}[\boldsymbol{\sigma}],
	\qquad
	\dot{\varepsilon}^{\text{vp}} =
	\begin{cases}
		0
		& \text{if } \tau_{\text{p}} < \sigma_0, \\
		a(\varepsilon - \varepsilon_0)^b \dot{\varepsilon}
		& \text{if } \tau_{\text{p}} \geq \sigma_0,
	\end{cases}
	\label{eq:eps_dot_vp}
\end{equation}
where $\tau_{\text{p}}=\|\operatorname{dev}[\boldsymbol{\sigma}]\|_F$, $a$,
$b$, and $\sigma_0$ are material constants, $\varepsilon_0$ marks the onset of
viscoplastic flow, $\dot{\varepsilon}$ is the scalar strain rate, and
$\varepsilon=\|\mathbf{E}\|_F$ is the norm of the Green-Lagrange strain
$\mathbf{E}=\tfrac{1}{2}(\mathbf{F}^T\mathbf{F}-\mathbf{I})$.
The evolution equations of $\bar{\mathbf{F}}^v$ and $\bar{\mathbf{F}}^{\text{vp}}$
and their exponential-map integration are given in~\ref{app:integration}.

The Helmholtz free energy is the sum of the equilibrium, non-equilibrium, and volumetric parts, degraded by the phase field $\phi\in[0,1]$,
\begin{equation}
	\rho_0\psi
	= g(\phi)\left[\psi_{\text{eq}}(\bar{\mathbf{B}}^{\text{ve}})
	+ \psi_{\text{neq}}(\bar{\mathbf{B}}^e)
	+ \psi_{\text{vol}}(J)\right].
	\label{eq:free_energy}
\end{equation}
Crack nucleation is governed by the stress-based driving force of
Eq.~\eqref{eq:energy_density}, so no split of the energy into tensile and
compressive parts is needed. The degradation function is
\begin{equation}
	g(\phi) = (1 - \phi)^2 + k,
	\label{eq:degradation_function}
\end{equation}
which satisfies $g(0)=1$, $g(1)=k$, and $g'(1)=0$ for $k\to0$, with a small
residual stiffness $k\ll1$ that avoids ill-conditioning at fully damaged
points~\cite{miehe2010thermodynamically}. Both the equilibrium and the
non-equilibrium energy are sums over $N_f$ fiber families,
\begin{equation}
	\rho_0\psi_{\text{eq}}
	= \sum_{i=1}^{N_f} v_f^{(i)}
	\!\left[\psi_{\text{eq}}^{\text{matrix}}(\bar{\mathbf{B}}^{\text{ve}})
	+ \psi_{\text{eq}}^{\text{fiber}(i)}
	(\bar{\mathbf{C}}^{\text{ve}}, \mathbf{a}_0^{(i)})\right],
	\label{eq:psi_eq}
\end{equation}
\begin{equation}
	\rho_0\psi_{\text{neq}}
	= \sum_{i=1}^{N_f} v_f^{(i)}
	\!\left[\psi_{\text{neq}}^{\text{matrix}}(\bar{\mathbf{B}}^e)
	+ \psi_{\text{neq}}^{\text{fiber}(i)}
	(\bar{\mathbf{C}}^e, \mathbf{a}_0^{(i)})\right],
	\label{eq:psi_neq}
\end{equation}
where $v_f^{(i)}$ and $\mathbf{a}_0^{(i)}$ are the volume fraction and the
reference direction of family $i$, and
$\bar{\mathbf{C}}^{\text{ve}}=(\bar{\mathbf{F}}^{\text{ve}})^T\bar{\mathbf{F}}^{\text{ve}}$
and $\bar{\mathbf{C}}^e=(\bar{\mathbf{F}}^e)^T\bar{\mathbf{F}}^e$ are the
right Cauchy-Green tensors. The fiber energy
$\psi^{\text{fiber}(i)}$ is the transversely isotropic energy
of~\cite{arash2019viscoelastic2}, given in~\ref{app:fiber}. The matrix
is neo-Hookean and its shear moduli follow a modified Kitagawa
law~\cite{unger2020effect},
\begin{equation}
	\psi_{\text{eq/neq}}^{\text{matrix}}
	= \tfrac{1}{2}\mu_{\text{eq/neq}}
	\!\left(\operatorname{tr}[\bar{\mathbf{B}}^{\text{ve/e}}] - 3\right),
	\qquad
	\mu_{\text{eq/neq}}(\theta)
	= \mu_{\text{eq/neq}}^0\bigl[2 - \exp(\alpha_{\theta}(\theta - \theta_0))\bigr],
	\label{eq:psi_matrix}
\end{equation}
where $\mu^0_{\text{eq}}$ and $\mu^0_{\text{neq}}$ are the shear moduli at the
reference temperature $\theta_0$. The volumetric energy is
\begin{equation}
	\rho_0\psi_{\text{vol}}
	= \tfrac{1}{2}K_v\!\left(\frac{J_m^2 - 1}{2} - \ln J_m\right),
	\label{eq:psi_vol}
\end{equation}
with the bulk modulus $K_v$ taken at the reference temperature.

The Cauchy stress follows from the free energy as
\begin{equation}
	\boldsymbol{\sigma} = g(\phi)\left(
	\boldsymbol{\sigma}_{\text{dev}} + \boldsymbol{\sigma}_{\text{vol}}\right),
	\label{eq:sigma_total}
\end{equation}
with the deviatoric and volumetric parts
\begin{equation}
	\boldsymbol{\sigma}_{\text{dev}}
	= J^{-1}\sum_{i=1}^{N_f} v_f^{(i)}
	\!\left[\boldsymbol{\sigma}_{\text{matrix}}
	+ \boldsymbol{\sigma}_{\text{fiber}}^{(i)}\right],
	\qquad
	\boldsymbol{\sigma}_{\text{vol}}
	= \tfrac{1}{2}K_v J^{-1}\!\left(J_m - \frac{1}{J_m}\right)\mathbf{I},
	\label{eq:sigma_dev}
\end{equation}
with the matrix stress
\begin{equation}
	\boldsymbol{\sigma}_{\text{matrix}}
	= \mu_{\text{eq}}\operatorname{dev}[\bar{\mathbf{B}}^{\text{ve}}]
	+ \mu_{\text{neq}}\operatorname{dev}[\bar{\mathbf{B}}^e].
	\label{eq:sigma_matrix}
\end{equation}
The fiber stress $\boldsymbol{\sigma}_{\text{fiber}}^{(i)}$ is given in~\ref{app:fiber}.

\subsection{Fiber orientation and fiber families}
\label{subsec:orientation}

The fiber orientation is described by the second-order orientation tensor
\begin{equation}
	\mathbf{A} = \frac{1}{N_{\text{fibers}}}
	\sum_{k=1}^{N_{\text{fibers}}} \mathbf{a}_k \otimes \mathbf{a}_k
	= \sum_{i=1}^{n_{\text{dim}}} \lambda_i \,
	\mathbf{n}_i \otimes \mathbf{n}_i,
	\label{eq:orientation_tensor}
\end{equation}
where $\mathbf{a}_k$ is the direction of fiber $k$, and $\lambda_i$ and
$\mathbf{n}_i$ are the eigenvalues and eigenvectors of $\mathbf{A}$, with
$\sum_i\lambda_i=1$. Each eigenvector defines a fiber family,
$\mathbf{a}_0^{(i)}=\mathbf{n}_i$, with the volume fraction
\begin{equation}
	v_f^{(i)} = v_f \cdot \frac{\lambda_i}{\sum_{j=1}^{N_f} \lambda_j},
	\qquad
	v_m = 1 - v_f,
	\label{eq:fiber_volume_fraction}
\end{equation}
where $v_f$ is the total fiber volume fraction and $v_m$ is the matrix volume
fraction. Thus $\sum_i v_f^{(i)}=v_f$. The tensor $\mathbf{A}$ determines the
fiber directions and weights in the free energy and, through $\hat{\mathbf{A}}$
below, the directional crack resistance.

\subsection{Anisotropic phase-field fracture model}
\label{sec:phase}

Let $\Omega_t\subset\mathbb{R}^{n_{\text{dim}}}$ be the deformed body with
boundary $\Gamma_t$. The strong form of the coupled problem for the
displacement $\mathbf{u}$ and the phase field $\phi$ reads
\begin{align}
	\nabla_{\mathbf{x}} \cdot \boldsymbol{\sigma} + \mathbf{b} &= \mathbf{0}
	\quad \text{in } \Omega_t, \qquad
	\boldsymbol{\sigma} \cdot \mathbf{n} = \bar{\mathbf{t}}
	\quad \text{on } \Gamma_t, \label{eq:momentum_spat}\\
	\frac{\mathcal{G}_c^\mathrm{eff}}{l_0}\phi
	- l_0 \nabla_{\mathbf{x}} \cdot \bigl(\mathcal{G}_c^\mathrm{eff}\,
	\hat{\mathbf{A}} \cdot \nabla_{\mathbf{x}}\phi\bigr) &= -g'(\phi)\mathcal{H}
	\quad \text{in } \Omega_t, \qquad
	\nabla_{\mathbf{x}}\phi \cdot \mathbf{n} = 0
	\quad \text{on } \Gamma_t, \label{eq:phasefield_spat}
\end{align}
where $\mathbf{b}$ is the body force, $\bar{\mathbf{t}}$ is the prescribed
traction, $l_0$ is the length scale, and $\mathcal{G}_c^\mathrm{eff}$ is the
effective critical energy release rate defined in Section~\ref{sec:fatigue}. In
the monotonic parent model $\mathcal{G}_c^\mathrm{eff}=\mathcal{G}_c$ is
constant. The fatigue extension replaces it with a spatially heterogeneous and
history-dependent field. The tensor
\begin{equation}
	\hat{\mathbf{A}} = \mathbf{I} + \hat{\alpha} \mathbf{A}
	\label{eq:aniso_tensor}
\end{equation}
introduces the directional dependence of the crack resistance, with the
dimensionless parameter $\hat{\alpha}\geq0$. Setting $\hat{\alpha}=0$ recovers
the isotropic model.

Irreversibility of crack growth is enforced by the history
field~\cite{miehe2010thermodynamically}
\begin{equation}
	\mathcal{H}(t) = \max_{\tau \in [0,t]} \mathcal{Y}(\tau),
	\label{eq:history_field}
\end{equation}
where the instantaneous driving force is controlled by the maximum principal
stress,
\begin{equation}
	\mathcal{Y}
	= \frac{\mathcal{G}_c^{\mathrm{eff}}}{l_0}\,
	\left\langle \frac{\sigma_I}{\sigma_d^{\mathrm{eff}}} - 1 \right\rangle_+^{2}.
	\label{eq:energy_density}
\end{equation}
Here $\langle\cdot\rangle_+=\max(\cdot,0)$ and $\sigma_I=\langle\sigma_1\rangle_+$,
where $\sigma_1$ is the largest principal value of the undegraded Cauchy stress,
that is, of $\boldsymbol{\sigma}$ in Eq.~\eqref{eq:sigma_total} with $g(\phi)=1$. The
effective toughness $\mathcal{G}_c^{\mathrm{eff}}$ and the effective crack
strength $\sigma_d^{\mathrm{eff}}$ are defined in Section~\ref{sec:fatigue}. No
driving force arises while $\sigma_I\le\sigma_d^{\mathrm{eff}}$. The
equilibrium, non-equilibrium, and volumetric contributions all enter through
the stress, so the stress carried by the viscoelastic branch also drives the
crack.

With admissible test functions $\boldsymbol{\eta}_u$ and $\eta_\phi$, the weak
form of Eqs.~\eqref{eq:momentum_spat} and~\eqref{eq:phasefield_spat} is
\begin{equation}
	\int_{\Omega_t} \boldsymbol{\sigma} : \nabla_{\mathbf{x}}\boldsymbol{\eta}_u
	\, dv
	- \int_{\Omega_t} \rho_t \mathbf{b} \cdot \boldsymbol{\eta}_u \, dv
	- \int_{\Gamma_t} \bar{\mathbf{t}} \cdot \boldsymbol{\eta}_u \, da
	= 0,
	\label{eq:weak_u_spat}
\end{equation}
\begin{equation}
	\int_{\Omega_t} \!\left[
	J^{-1}g'(\phi)\mathcal{H}\,\eta_\phi
	+ J^{-1}\frac{\mathcal{G}_c^\mathrm{eff}}{l_0}\phi\,\eta_\phi
	+ J^{-1}\mathcal{G}_c^\mathrm{eff} l_0
	\nabla_{\mathbf{x}}\phi \cdot \hat{\mathbf{A}} \cdot \nabla_{\mathbf{x}}\eta_\phi
	\right]\! dv = 0.
	\label{eq:weak_phi_spat}
\end{equation}
The equations are discretized with isoparametric Q4 elements and $2\times2$
Gauss quadrature for both fields and solved with a staggered
algorithm~\cite{miehe2010phase}, in which the displacement and phase-field
problems are solved in turn within each load step by Newton-Raphson iteration.
The internal variables are updated by the exponential-map integrator of~\ref{app:integration}. The consistent tangent of the displacement
problem is computed numerically by perturbing the deformation gradient, as
described in~\ref{app:fem}.

\section{Phase-field fatigue fracture model}
\label{sec:fatigue}

Under cyclic loading, the gradual subcritical damage of the matrix is driven by two physically different mechanisms. If $\tau_p > \sigma_0$ during the loading half-cycle, irreversible plastic flow builds up a permanent deformation of the matrix. This mechanism dominates fatigue at low and moderate cycle numbers, where the stress amplitude pushes the material beyond yield. Below $\sigma_0$, the non-equilibrium polymer network is still stretched and only partly recovered in each cycle. The repeated positive increments of $\Psi^\mathrm{neq}$ cause progressive chain scission and microcracking. This mechanism controls HCF at sub-yield amplitudes, for which the purely viscoplastic accumulator gives no damage. Both mechanisms arise naturally from the free energy structure of the existing VE-VP model. Apart from the two fatigue history variables defined below, no additional constitutive assumptions are needed.

The fatigue effects are introduced into the anisotropic phase-field model of Section~\ref{sec:model} solely through the scalar fields $\mathcal{G}_c^\mathrm{eff}$ and $\sigma_d^\mathrm{eff}$. As a result, the model keeps both the anisotropic stress response of the fiber families and the directional crack resistance given by $\hat{\mathbf{A}}$. The former also governs the evolution of the fatigue history variables. Because $\mathcal{G}_c^\mathrm{eff}$ scales both terms of the regularized crack surface energy in Eq.~\eqref{eq:phasefield_spat}, fatigue reduces the crack resistance by an equal factor in all directions. The properties that ensure the consistency of the formulation and the control of the cycle-jump scheme are given as remarks in the subsections below.

\subsection{Fatigue history variables}
\label{sec:fatigue_vars}

Fatigue damage accumulation is tracked through two scalar history
variables, each obtained by integrating a Heaviside-filtered rate over
the loading history so that only loading increments, not unloading or
elastic reversals, contribute.

The accumulated viscoplastic strain $\kappa(\mathbf{x},t)$ is dimensionless. It grows whenever the viscoplastic yield condition is active and the material is loading,
\begin{equation}
	\kappa(\mathbf{x},t) = \int_0^t H(\tau_p - \sigma_0)\,H(\dot{\varepsilon}^{\text{vp}})\,
	\dot{\varepsilon}^{\text{vp}}\,\mathrm{d}s,
	\label{eq:kappa_def}
\end{equation}
where $\dot{\varepsilon}^{\text{vp}}$ and $\dot{\varepsilon}$ are the viscoplastic and the total strain rates of Eq.~\eqref{eq:eps_dot_vp}, and $H(\cdot)$ is the Heaviside step function. The two filters ensure that $\kappa$ accumulates only above the
yield threshold $\sigma_0$ and only while the material is actively
loading, so it stays constant during unloading and during purely
sub-yield deformation.

A second accumulator, $\bar{\kappa}(\mathbf{x},t)$, tracks
sub-yield viscoelastic damage that $\kappa$ cannot capture, since
HCF failure often occurs at stress levels below
$\sigma_0$. It integrates the positive part of the rate of the
non-equilibrium stored strain-energy density $\Psi^{\mathrm{neq}}=\rho_0\psi_\mathrm{neq}$ of
Eq.~\eqref{eq:psi_neq}, evaluated without degradation, that is, with $g(\phi)=1$,
\begin{equation}
	\bar{\kappa}(\mathbf{x},t) = \int_0^t H(\dot{\Psi}^{\mathrm{neq}})\,
	\dot{\Psi}^{\mathrm{neq}}\,\mathrm{d}s,
	\label{eq:kappa_bar_def}
\end{equation}
with units of MPa. In the numerical implementation $\dot{\Psi}^{\mathrm{neq}}$
is not evaluated from a closed-form rate law. It is obtained as the
increment of the converged $\Psi^{\mathrm{neq}}$ between
successive load steps, and only positive increments (loading) are
accumulated, consistent with the discrete-increment treatment used
elsewhere in this section for rates of accumulated state variables. The accumulator $\bar\kappa$ therefore measures the energy cycled through the non-equilibrium network and not a dissipated energy, although this stored energy is what the viscous flow relaxes and dissipates.

The two accumulators combine into a single fatigue driving force,
\begin{equation}
	\mathcal{D}_{\mathrm{fat}} = \kappa + \beta\bar{\kappa},
	\label{eq:D_fat}
\end{equation}
where the coupling weight $\beta$ (units MPa$^{-1}$, since $\kappa$ is
dimensionless and $\bar\kappa$ carries units of MPa) sets the relative
contribution of viscoelastic to viscoplastic damage, and is calibrated
against experimental fatigue data. This driving force degrades the
fracture toughness through
\begin{equation}
	\mathcal{G}_c^{\mathrm{eff}} = \mathcal{G}_c\,\mathcal{F}(\mathcal{D}_{\mathrm{fat}}),
	\label{eq:Gc_eff}
\end{equation}
\begin{equation}
	\mathcal{F}(\mathcal{D}_{\mathrm{fat}}) =
	\exp\!\left(-\left\langle \frac{\mathcal{D}_{\mathrm{fat}} - \mathcal{D}_T}
	{\mathcal{D}_c}\right\rangle\right),
	\label{eq:F_deg}
\end{equation}
where $\langle\cdot\rangle$ denotes the Macaulay bracket, $\mathcal{D}_T$
is a threshold below which no degradation occurs, and $\mathcal{D}_c$
sets the rate of toughness decay once the threshold is exceeded.

The threshold $\mathcal{D}_T$ defines the local activation level, and the toughness remains intact until $\mathcal{D}_{\mathrm{fat}}$ exceeds it. Beyond this threshold, $\mathcal{D}_c$ controls how fast $\mathcal{G}_c^{\mathrm{eff}}$ decreases for a given increase in $\mathcal{D}_{\mathrm{fat}}$. The weight $\beta$ decides whether sub-yield viscoelastic damage enters $\mathcal{D}_{\mathrm{fat}}$. For $\beta = 0$, only $\kappa$ is active. Because $\kappa$ grows only when $\tau_p$ exceeds the yield threshold $\sigma_0$, the model is then limited to the low-cycle regime above yield. A nonzero $\beta$ allows $\bar\kappa$ to drive the degradation also in sub-yield states, which extends the model to HCF. This argument is made precise below.

At finite deformation, the fatigue variables must be independent of the observer. Consider a superposed rotation $\mathbf{F}\to\mathbf{Q}(t)\mathbf{F}$, with $\bar{\mathbf{F}}^v$ and $\bar{\mathbf{F}}^{\mathrm{vp}}$ unchanged as required for objective flow rules. The stress then transforms as $\boldsymbol{\sigma}\to\mathbf{Q}\boldsymbol{\sigma}\mathbf{Q}^T$, while the Green-Lagrange strain remains the same. Consequently, $\tau_p$, $\varepsilon$, $\dot\varepsilon$, $\dot\varepsilon^{\mathrm{vp}}$, and the principal stress $\sigma_I$ are not affected. The elastic gradient transforms as $\bar{\mathbf{F}}^e\to\mathbf{Q}\bar{\mathbf{F}}^e$. Hence $\operatorname{tr}\bar{\mathbf{B}}^e$, the fiber invariants $\bar I_4^{(i)}$ and $\bar I_5^{(i)}$ (\ref{app:fiber}), and $\Psi^\mathrm{neq}$ are also unchanged. It follows that the integrands of Eqs.~\eqref{eq:kappa_def} and~\eqref{eq:kappa_bar_def} are invariant. The same holds for $\kappa$, $\bar\kappa$, $\mathcal{D}_\mathrm{fat}$, $\mathcal{G}_c^\mathrm{eff}$, and the crack driving force $\mathcal{Y}$ of Eq.~\eqref{eq:energy_density}.

The two accumulators describe two distinct fatigue mechanisms. The viscoplastic accumulator $\kappa$ increases only where $\tau_p>\sigma_0$. If a material point satisfies $\tau_p\le\sigma_0$ throughout the loading, Eq.~\eqref{eq:kappa_def} yields $\kappa=0$. For $\beta=0$, no fatigue damage therefore develops below yield. In contrast, the viscoelastic accumulator $\bar\kappa$ increases in every cycle in which $\Psi^\mathrm{neq}$ changes, regardless of the stress level. For $\beta>0$, this increase contributes to $\mathcal{D}_\mathrm{fat}$. During one cycle $[t_0,t_0+T]$, $\bar\kappa$ collects the total rise of $\Psi^\mathrm{neq}$. In a periodic state, this rise is independent of the starting point of the cycle. If the cycle starts at the minimum, $\Psi^\mathrm{neq}$ has to rise to its maximum, which gives
\begin{equation}
	\Delta\bar\kappa=\int_{t_0}^{t_0+T} H(\dot{\Psi}^{\mathrm{neq}})\,
	\dot{\Psi}^{\mathrm{neq}}\,\mathrm{d}s
	\;\ge\;\Psi^\mathrm{neq}_{\max}-\Psi^\mathrm{neq}_{\min}.
	\label{eq:kbar_cycle}
\end{equation}
Equality holds when $\Psi^\mathrm{neq}$ has a single maximum and a single minimum per cycle. Every sub-yield cycle with a nonzero range of $\Psi^\mathrm{neq}$ therefore yields a positive increment of $\bar\kappa$, and this extends the model to HCF. A hold at fixed displacement adds nothing to $\bar\kappa$, since $\Psi^\mathrm{neq}$ can then only decrease through relaxation.

\subsection{Adaptive cycle-jump strategy}
\label{sec:cycle_jump}

Simulating every fatigue cycle explicitly is computationally prohibitive, so an adaptive cycle-jump strategy is adopted. The fatigue accumulators $\kappa$ and $\bar{\kappa}$ are extrapolated over blocks of skipped cycles. The mechanical internal variables $\bar{\mathbf{F}}^v$ and $\bar{\mathbf{F}}^{\mathrm{vp}}$ are updated only within the explicitly simulated blocks. These variables are not extrapolated, because a cycle-consistent extrapolation of them is not well defined for a VE-VP constitutive history. Linear extrapolation is applied only to the scalar accumulators $\kappa$ and $\bar{\kappa}$. Since the mechanical state is frozen during a jump, the measured rates are corrected for the age of the state through a damage clock. The crack strength is corrected in the same manner. Both corrections are derived below from the reaction force history. The same rules apply to LCF and HCF. Two cycle counters are introduced. $N$ denotes the number of elapsed cycles, and $n\le N$ denotes the number of explicitly simulated cycles.

\begin{figure}[H]
	\centering
	\includegraphics[scale=0.9]{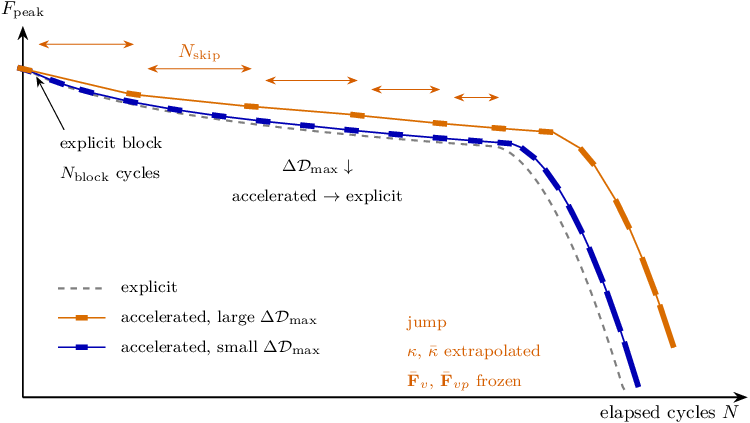}
	\caption{Schematic of the adaptive cycle-jump strategy in the peak force $F_{\mathrm{peak}}$ against the elapsed cycles $N$. The gray dashed curve is the explicit simulation. The short thick segments are the explicitly simulated blocks of $N_{\mathrm{block}}$ cycles. Between them the thin lines are the jumps, in which $\kappa$ and $\bar{\kappa}$ are extrapolated over $N_{\mathrm{skip}}$ cycles while the mechanical internal variables are frozen. A large $\Delta\mathcal{D}_{\max}$ allows long jumps and can overestimate the force, and the accelerated curve approaches the explicit one as $\Delta\mathcal{D}_{\max}$ decreases.}
	\label{fig:cycle_jump_schematic}
\end{figure}

Fig.~\ref{fig:cycle_jump_schematic} illustrates the scheme in terms of the peak force $F_\mathrm{peak}$ versus the elapsed cycles $N$. The gray dashed curve represents the explicit simulation, in which every cycle is computed. The accelerated simulation switches between two phases. During an explicit block of $N_\mathrm{block}$ cycles, shown as a short thick segment, all fields are advanced. During a jump over $N_\mathrm{skip}$ cycles, shown as a thin line, only $\kappa$ and $\bar\kappa$ are extrapolated, and $\bar{\mathbf{F}}^v$ and $\bar{\mathbf{F}}^{\mathrm{vp}}$ are held fixed. As a result, the accelerated curve departs from the explicit one and may overestimate the force after a jump. This deviation is controlled by the bound $\Delta\mathcal{D}_{\max}$ on the toughness drop per jump, which is defined below. A large bound (orange) permits long jumps and produces a visible deviation. A small bound (blue) shortens the jumps and brings the accelerated curve closer to the explicit one. This is confirmed by the analysis of the limiter below and by the convergence study in Section~\ref{sec:convergence}.

The block length $N_{\mathrm{block}}$, defined as the number of explicitly simulated cycles per block, has to be sufficient for the internal variables to approach the periodic state of the loading cycle. Only then does the damage rate measured in the block reflect the current stage of the fatigue process. Departures from the periodic state decay through viscous flow. The decay rate depends exponentially on the local stress, so a single relaxation time cannot describe the entire domain. For this reason, the block length is determined by a direct check on the damage increment per cycle.

The increment of $\mathcal{D}_{\mathrm{fat}}$ at integration point $\mathbf{x}$ during cycle $k$ of the block is written as $\Delta\mathcal{D}_k(\mathbf{x})=\Delta\kappa_k(\mathbf{x})+\beta\,\Delta\bar{\kappa}_k(\mathbf{x})$. Here $\Delta\kappa_k$ and $\Delta\bar{\kappa}_k$ are the increments of $\kappa$ and $\bar{\kappa}$ within that cycle. With the peak increment $\delta_k=\max_{\mathbf{x}}\Delta\mathcal{D}_k(\mathbf{x})$, the relative change between two successive cycles reads
\begin{equation}
	\rho_k=\frac{\left|\delta_k-\delta_{k-1}\right|}{\delta_k},\qquad k\ge 2 .
	\label{eq:rho_k}
\end{equation}
A pointwise criterion is avoided. Individual points may gain or lose one load step of the interval in which $\tau_p>\sigma_0$, and this keeps such a measure close to 0.1 in every cycle. The block ends at the first cycle in which the peak increment becomes stationary,
\begin{equation}
	N_{\mathrm{block}}=\min\left\{\,k\in\{2,\dots,N_{\mathrm{block}}^{\max}\}\;\middle|\;\rho_k\le\varepsilon_\rho\right\},
	\label{eq:Nblock}
\end{equation}
and $N_{\mathrm{block}}=N_{\mathrm{block}}^{\max}$ is used if no such $k$ is found. In this case, no cycles are skipped after the block, and the simulation proceeds explicitly with the next block. The tolerance and the upper bound are set to $\varepsilon_\rho=0.01$ and $N_{\mathrm{block}}^{\max}=10$, respectively.


After the block has ended, the per-cycle rate is evaluated over the second half of the explicit block,
\begin{equation}
	\dot{\kappa}_\mathrm{cyc}(\mathbf{x})
	= \frac{\kappa(\mathbf{x},\,N+N_\mathrm{block})
		- \kappa(\mathbf{x},\,N+\lfloor N_\mathrm{block}/2\rfloor)}
	{\lceil N_\mathrm{block}/2 \rceil},
	\label{eq:kappa_rate}
\end{equation}
where $\lfloor\cdot\rfloor$ and $\lceil\cdot\rceil$ are the floor and ceiling operators. They round their argument down and up to the nearest integer, respectively. The denominator thus equals the number of cycles between the two evaluation points, also for odd $N_\mathrm{block}$. The rate $\dot{\bar{\kappa}}_\mathrm{cyc}$ is computed in the same way. In this manner, the transient cycles at the beginning of the block are excluded, and only the stabilized part is used for the extrapolation.

The mechanical state is held fixed during a jump. At the elapsed cycle $N$, the state therefore corresponds to the one obtained after $n<N$ explicit cycles. Under displacement control, the reaction force decreases slowly with the number of cycles. The frozen state thus carries a higher stress than the aged material would carry at cycle $N$, and the rates measured in the block are overestimated. The age of the state is estimated from the normalized peak force $f_j=F_j/F_\mathrm{peak}$ of the explicit blocks. Over the uncracked blocks within the window $n/2\le n_j\le n$, the force is described by
\begin{equation}
	L(n)=f_0+s\ln n,\qquad s\le 0 ,
	\label{eq:force_line}
\end{equation}
in which $f_0$ and $s$ follow from a least-squares fit. Blocks with $\phi_{\max}>10^{-3}$ are excluded, so that the force drop due to cracking is not extrapolated. The fit is applied once $n\ge 40$ and the window contains at least five blocks. The age ratio is defined as
\begin{equation}
	R=\min\left\{1,\;\max\left\{R_{\min},\;\frac{L(N)}{L(n)}\right\}\right\}.
	\label{eq:R_age}
\end{equation}

It is equal to one when no cycles have been skipped and decreases as the elapsed cycle count moves ahead of the simulated one. Its lower bound is set to $R_{\min}=0.2$. The rates are then scaled by the damage clock
\begin{equation}
	\dot{\kappa}_\mathrm{eff}=c\,\dot{\kappa}_\mathrm{cyc},\qquad
	\dot{\bar{\kappa}}_\mathrm{eff}=\bar{c}\,\dot{\bar{\kappa}}_\mathrm{cyc},\qquad
	c=R^{m_\kappa},\qquad \bar{c}=R^{m_{\bar\kappa}}.
	\label{eq:clock}
\end{equation}

The scaling applies only to the extrapolated increment. Since the cycles of the explicit block are computed at the true mechanical state, their increments of $\kappa$ and $\bar{\kappa}$ are retained unchanged. The exponents $m_\kappa$ and $m_{\bar\kappa}$ link the damage rates to the normalized force through $\dot{\kappa}_\mathrm{cyc}\propto f^{m_\kappa}$ and $\dot{\bar{\kappa}}_\mathrm{cyc}\propto f^{m_{\bar\kappa}}$. A fit to the rates of explicit simulations gave $m_\kappa=6$ and $m_{\bar\kappa}=1.5$. These values are fixed once and applied to all load levels. Like $R_{\min}$, they are numerical parameters.

The same age ratio corrects the crack criterion. The crack driving force is evaluated on the frozen state, whose stress is too high by the factor $1/R$. The crack strength is therefore raised accordingly,
\begin{equation}
	\sigma_{d}^\mathrm{eff}=\sigma_{d}\,\frac{\max\{\mathcal{F},\,10^{-3}\}}{R},
	\label{eq:sigma_d_eff}
\end{equation}
where $\mathcal{F}=\mathcal{G}_c^\mathrm{eff}/\mathcal{G}_c$ is the toughness fraction, which is bounded below by $10^{-3}$ so that the crack strength stays positive. The ratio $R$ is evaluated from the most recent force line, at the beginning of the block for Eq.~\eqref{eq:sigma_d_eff} and at the end of the block for Eq.~\eqref{eq:clock}.

The fatigue variables are extrapolated forward over $N_\mathrm{skip}$ cycles,
\begin{equation}
	\kappa(\mathbf{x}) \;\leftarrow\; \kappa(\mathbf{x})
	+ N_\mathrm{skip}\,\dot{\kappa}_\mathrm{eff}(\mathbf{x}),
	\qquad
	\bar{\kappa}(\mathbf{x}) \;\leftarrow\; \bar{\kappa}(\mathbf{x})
	+ N_\mathrm{skip}\,\dot{\bar{\kappa}}_\mathrm{eff}(\mathbf{x}).
	\label{eq:extrapolation}
\end{equation}
%
%

Once the extrapolation is complete, $\mathcal{G}_c^\mathrm{eff}(\mathbf{x})$ is updated through Eqs.~(\ref{eq:D_fat}) and (\ref{eq:Gc_eff}). A single phase-field solve with frozen mechanical fields then advances $\phi$. The extrapolation cannot reverse the fatigue damage, as shown next. The integrands of Eqs.~\eqref{eq:kappa_def} and~\eqref{eq:kappa_bar_def} are nonnegative, so $\kappa$ and $\bar\kappa$ cannot decrease within an explicit block. The rates of Eq.~\eqref{eq:kappa_rate} are differences of nondecreasing values and are therefore nonnegative as well. Because $R\in[R_{\min},1]$ with $R_{\min}>0$, the clocks $c=R^{m_\kappa}$ and $\bar c=R^{m_{\bar\kappa}}$ take values in $(0,1]$. For $N_\mathrm{skip}\ge0$, every term added in Eq.~\eqref{eq:extrapolation} is thus nonnegative. It follows that $\kappa$, $\bar\kappa$, and $\mathcal{D}_\mathrm{fat}$ never decrease, neither during explicit blocks nor during jumps. The function $\mathcal{F}$ in Eq.~\eqref{eq:F_deg} does not increase with $\mathcal{D}_\mathrm{fat}$ and takes values in $(0,1]$. Hence the toughness $\mathcal{G}_c^\mathrm{eff}=\mathcal{G}_c\mathcal{F}$ can only decrease or stay constant. This argument concerns the degradation of the toughness only. The irreversibility of the crack itself is ensured separately by the history field $\mathcal{H}$ and the constraint $\dot\phi\ge0$. In general, the crack strength $\sigma_d^\mathrm{eff}$ is not monotone, since $R$ may increase or decrease from one block to the next.

The skip size $N_\mathrm{skip}$ is first estimated from the per-cycle
fatigue driving-force rate,
\begin{equation}
	N_\mathrm{skip}
	= \min\!\Biggl(
	\alpha\,
	\frac{\mathcal{D}_c}{\dot{\mathcal{D}}_\mathrm{fat,max}},\;
	N_\mathrm{skip}^\mathrm{max}
	\Biggr),
	\label{eq:N_skip}
\end{equation}
where
$\dot{\mathcal{D}}_\mathrm{fat,max}
= \max_{\mathbf{x} \in \Omega}
(\dot{\kappa}_\mathrm{eff} + \beta\,\dot{\bar{\kappa}}_\mathrm{eff})$
is taken over the whole domain $\Omega$ rather than an active process zone,
since $\dot{\kappa}_\mathrm{eff}$ and $\dot{\bar{\kappa}}_\mathrm{eff}$ are
negligible outside the process zone by construction, so the two are
effectively equivalent in practice. Here $\alpha = 0.3$ is a safety factor,
and $N_\mathrm{skip}^\mathrm{max}=10^4$ is an upper bound.

This estimate is then corrected so that the resulting drop in the
effective toughness fraction stays bounded. The trial fatigue driving
force after $N_\mathrm{skip}$ cycles is $\mathcal{D}_\mathrm{fat}^\mathrm{trial}
= \mathcal{D}_\mathrm{fat} + N_\mathrm{skip}\dot{\mathcal{D}}_\mathrm{fat}$, with the rate
$\dot{\mathcal{D}}_\mathrm{fat}=\dot{\kappa}_\mathrm{eff}+\beta\,\dot{\bar{\kappa}}_\mathrm{eff}$ at each point, giving the toughness fractions
$\mathcal{F} = \mathcal{F}(\mathcal{D}_\mathrm{fat})$ and
$\mathcal{F}^\mathrm{trial} = \mathcal{F}(\mathcal{D}_\mathrm{fat}^\mathrm{trial})$
from Eq.~\eqref{eq:F_deg}. If the projected drop exceeds a prescribed bound,
\begin{equation}
	\mathcal{F} - \mathcal{F}^\mathrm{trial} > \Delta\mathcal{D}_\mathrm{max},
	\label{eq:Df_limiter}
\end{equation}
$N_\mathrm{skip}$ is reduced,
\begin{equation}
	N_\mathrm{skip}^{\mathrm{lim}}(\mathbf{x})
	=\frac{\mathcal{D}_T-\mathcal{D}_\mathrm{fat}(\mathbf{x})-\mathcal{D}_c\ln\!\left(\mathcal{F}(\mathbf{x})-\Delta\mathcal{D}_\mathrm{max}\right)}{\dot{\mathcal{D}}_\mathrm{fat}(\mathbf{x})},
	\label{eq:Nskip_lim}
\end{equation}
\begin{equation}
	N_\mathrm{skip}\;\leftarrow\;
	\min\!\left(N_\mathrm{skip},\;
	\left\lfloor \min_{\mathbf{x}\in\Omega_{\mathcal{F}}} N_\mathrm{skip}^{\mathrm{lim}}(\mathbf{x})\right\rfloor\right),
	\label{eq:Nskip_correction}
\end{equation}
where the set of points that can limit the jump is
\begin{equation}
	\Omega_{\mathcal{F}}=\left\{\mathbf{x}\;\middle|\;\mathcal{F}(\mathbf{x})>\Delta\mathcal{D}_\mathrm{max},\;\dot{\mathcal{D}}_\mathrm{fat}(\mathbf{x})>0\right\}.
	\label{eq:Omega_F}
\end{equation}

The bound $\Delta\mathcal{D}_\mathrm{max}$ is the numerical parameter that governs the accuracy of the scheme, and its meaning can be stated exactly. At a given point, a jump of $N_\mathrm{skip}$ cycles reduces the toughness fraction by $\mathcal{F}-\mathcal{F}_{N_\mathrm{skip}}\le\Delta\mathcal{D}_\mathrm{max}$ if and only if $N_\mathrm{skip}\le N_\mathrm{skip}^\mathrm{lim}$ of Eq.~\eqref{eq:Nskip_lim}. The proof is given in \ref{app:limiter}, Eq.~\eqref{eq:Nlim_iff}. Consequently, every jump keeps the drop of $\mathcal{F}$ below $\Delta\mathcal{D}_\mathrm{max}$ at all points. This also holds for jumps shortened by Eqs.~\eqref{eq:Nskip_correction} and~\eqref{eq:Nskip_cap}. Close to a crack the rates are high, which gives $N_\mathrm{skip}^\mathrm{lim}<1$, and no jump is performed. The scheme thus falls back to the explicit simulation wherever resolution is required. Since $N_\mathrm{skip}^\mathrm{lim}$ is proportional to $\Delta\mathcal{D}_\mathrm{max}$ (Eq.~\eqref{eq:Nlim_approx}), the scheme also approaches the explicit simulation everywhere as $\Delta\mathcal{D}_\mathrm{max}\to0$.

Two additional limits are imposed. No jump is permitted before $N$ reaches $N_\mathrm{jump}^{\min}=40$ cycles, because the rates of the first cycles are not yet representative of the later process. Beyond this point, the jump length is restricted to a fraction of the elapsed cycles,
\begin{equation}
	N_\mathrm{skip}\;\leftarrow\;\min\left(N_\mathrm{skip},\;\lfloor \chi N\rfloor\right),\qquad \chi=0.1.
	\label{eq:Nskip_cap}
\end{equation}

This restriction ensures that the age ratio $R$ changes only slightly during one jump. The limit of Eq.~\eqref{eq:Nskip_cap} increases with $N$ and controls the early stage of the fatigue life. In contrast, the limit of Eq.~\eqref{eq:Nskip_correction} is almost independent of $N$ and controls the later stage. There, it scales with $\Delta\mathcal{D}_\mathrm{max}$ as given by Eq.~\eqref{eq:Nlim_closed}. Finally, no cycles are skipped when the rate has not become stationary by the end of the block. This is the case when $N_\mathrm{block}=N_\mathrm{block}^{\max}$ is reached with $\rho_k>\varepsilon_\rho$. The cycle counters are advanced as
\begin{equation}
	N \;\leftarrow\; N + N_\mathrm{block} + N_\mathrm{skip},\qquad
	n \;\leftarrow\; n + N_\mathrm{block}.
	\label{eq:N_update}
\end{equation}
Algorithm~\ref{alg:cycle_jump} summarizes the complete procedure. It follows the sequence shown in Fig.~\ref{fig:cycle_jump_schematic}, in which each pass of the loop is one explicit block followed by one jump.

\begin{algorithm}[h!]
	\caption{Adaptive cycle-jump for VE-VP phase-field fatigue}
	\label{alg:cycle_jump}
	\begin{algorithmic}[1]
		\State \textbf{Initialize}
		$\kappa \leftarrow \mathbf{0}$,\;
		$\bar{\kappa} \leftarrow \mathbf{0}$,\;
		$\mathcal{G}_c^\mathrm{eff} \leftarrow \mathcal{G}_c$,\;
		$N \leftarrow 0$,\;
		$n \leftarrow 0$,\;
		$F_\mathrm{peak} \leftarrow 0$,\;
		$R \leftarrow 1$
		\Repeat
		\State Compute $\sigma_d^\mathrm{eff}$ via Eq.~(\ref{eq:sigma_d_eff})
		with the current $R$
		\State \textbf{Explicit block} simulate cycles using the staggered VE-VP solver
		($\mathbf{u}$-solve $\to$ $\phi$-solve) until $\rho_k\le\varepsilon_\rho$
		or $N_\mathrm{block}^{\max}$ is reached,
		Eqs.~(\ref{eq:rho_k}) and (\ref{eq:Nblock});
		track $F_\mathrm{block}$
		\State Accumulate $\kappa$, $\bar{\kappa}$ at all Gauss points
		via Eqs.~(\ref{eq:kappa_def}) and (\ref{eq:kappa_bar_def})
		\State Update $F_\mathrm{peak} \leftarrow
		\max(F_\mathrm{peak},\,F_\mathrm{block})$,
		store $(n,\,F_\mathrm{block}/F_\mathrm{peak},\,\phi_{\max})$
		and refit the force line, Eq.~(\ref{eq:force_line})
		\State Measure per-cycle rates over second half of block
		via Eq.~(\ref{eq:kappa_rate})
		\State \textbf{Damage clock} compute $R$, $c$, $\bar c$
		via Eqs.~(\ref{eq:R_age}) and (\ref{eq:clock}),
		scale the rates
		\State Estimate $N_\mathrm{skip}$ via Eq.~(\ref{eq:N_skip});
		reduce it via Eqs.~(\ref{eq:Nskip_correction}) and (\ref{eq:Nskip_cap});
		set $N_\mathrm{skip}\leftarrow 0$ if $N<N_\mathrm{jump}^{\min}$
		or the rate is not stationary
		\State \textbf{Extrapolation} advance $\kappa$, $\bar{\kappa}$
		by $N_\mathrm{skip}$ cycles via Eq.~(\ref{eq:extrapolation})
		\State \textbf{Toughness update} compute $\mathcal{D}_\mathrm{fat}$,
		$\mathcal{F}$, $\mathcal{G}_c^\mathrm{eff}$
		via Eqs.~(\ref{eq:D_fat}) and (\ref{eq:Gc_eff})
		\State \textbf{Phase-field solve} solve Eq.~(\ref{eq:phasefield_spat})
		with updated $\mathcal{G}_c^\mathrm{eff}$; enforce $\dot{\phi}\geq 0$
		\State Advance $N$ and $n$ via Eq.~(\ref{eq:N_update})
		\Until{$F_\mathrm{block} < f_\mathrm{drop}\cdot F_\mathrm{peak}$
			\textbf{ or } $N \geq N_f^\mathrm{max}$}
		\State \textbf{Output} $\phi(\mathbf{x},N)$,\;
		$\kappa(\mathbf{x},N)$,\; $\bar{\kappa}(\mathbf{x},N)$,\;
		cycles to failure $N_f$
	\end{algorithmic}
\end{algorithm}

The fatigue model is based on three assumptions. (1) The fatigue parameters $\mathcal{D}_T$, $\mathcal{D}_c$, and $\beta$ were calibrated and validated with experimental fatigue data for the CF/epoxy system considered here (Section~\ref{sec:valid}). Their individual effects on the fatigue life and their interactions are quantified by a full factorial sensitivity analysis (Section~\ref{sec:sensitivity_analysis}). The cycle-jump parameters $\Delta\mathcal{D}_\mathrm{max}$, $\varepsilon_\rho$, $N_\mathrm{block}^{\max}$, $\alpha$, $\chi$, $N_\mathrm{jump}^{\min}$, $R_{\min}$, $m_\kappa$, and $m_{\bar\kappa}$ are purely numerical. They were chosen by comparison with the explicit simulation and play no role in the material calibration. The influence of $\Delta\mathcal{D}_\mathrm{max}$ on the predicted life is examined in Section~\ref{sec:convergence}. (2) The extrapolation of Eq.~\eqref{eq:extrapolation} relies on the per-cycle rates $\dot\kappa_\mathrm{cyc}$ and $\dot{\bar\kappa}_\mathrm{cyc}$, which are measured once in the explicit block. This is justified only when the loading amplitude during the skipped cycles equals that of the block in which the rates were measured. The condition is satisfied for the constant-amplitude loading studied here. For variable-amplitude loading, the rates would have to be evaluated again for every amplitude level, and the force line of Eq.~\eqref{eq:force_line} would need to be refitted. This extension is left for future work. (3) The internal variables $\bar{\mathbf{F}}^v$ and $\bar{\mathbf{F}}^{\mathrm{vp}}$ remain frozen during the skip phase. The damage clock and the crack strength compensate for this effect on the accumulated damage and on crack initiation. Their accuracy, however, is limited by that of the force line, which is extrapolated from the explicit cycles $n$ to the elapsed cycles $N$.

In every explicit block, each integration step obeys the dissipative Argon flow rule and the threshold-based viscoplastic flow rule of Section~\ref{sec:model}. The dissipation inequality of the VE-VP model is therefore satisfied within the blocks. During the skipped cycles, only the scalar accumulators $\kappa$ and $\bar\kappa$ are advanced, and, as shown above, this cannot reverse the fatigue damage. The approximation across the skip boundaries concerns the mechanical trajectory. The dissipation inequality is not evaluated across the jumps.

\section{Results and discussion}
\label{sec:results}


This section reports four groups of numerical results. In Section~\ref{sec:valid}, the viscoelastic, viscoplastic, and fatigue parameters are calibrated with monotonic and fatigue experiments on a short carbon fiber-reinforced epoxy dog-bone specimen. The calibrated fatigue parameters are then validated against an experimental point that was excluded from the fit. Section~\ref{sec:convergence} evaluates the accuracy and the speedup of the cycle-jump method for decreasing values of the jump bound $\Delta\mathcal{D}_\mathrm{max}$. In Section~\ref{sec:sensitivity_analysis}, the influence of the fatigue parameters $\mathcal{D}_T$, $\mathcal{D}_c$, and $\beta$ on the fatigue life is quantified by a variance decomposition of $\ln N_f$ based on a full factorial study. This analysis also shows how their contributions change from the LCF to the HCF regime. Finally, Section~\ref{sec:sen_simulation} applies the calibrated model to a single-edge notched specimen. It examines the effects of loading amplitude, temperature, fiber content, and fiber orientation on fatigue crack initiation and growth.

\subsection{Calibration and validation}
\label{sec:valid}

Table~\ref{tab:material_parameters} summarizes the viscoelastic, viscoplastic, phase-field, fatigue, and cycle-jump parameters adopted in this section, grouped by the mechanism they control. The phase-field parameters $\mathcal{G}_c$, $l_0$, and $\hat{\alpha}$ are adopted directly from our earlier studies on a comparable short fiber-reinforced polymer system~\citep{arash2026phase,ARASH2027114194}. They are not recalibrated in this work. In particular, $\mathcal{G}_c$ is assumed constant and independent of the fiber content. This simplification is necessary because no experimental data linking the fracture toughness to the fiber content are available for this system.

\begin{figure}[H]
	\centering
	{\includegraphics[scale=0.8]{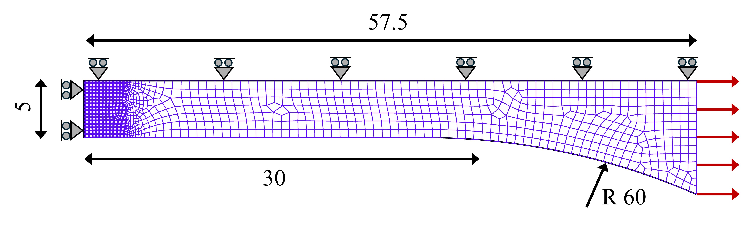}}
	\caption{Dimensions, loading, and boundary conditions imposed on a quarter of the dog-bone specimen, with a thickness of 2.3 mm. All dimensions are in millimeters.}
	\label{fig:db_mesh}
\end{figure}

In the first step, a single-objective genetic algorithm was used to identify the viscoelastic and viscoplastic parameters of Table~\ref{tab:material_parameters} by fitting the entire experimental stress-strain curve of~\cite{hematipour2025progressive}. The calibration was based on the monotonic tensile response of a short carbon fiber-reinforced epoxy dog-bone specimen with 5~wt\% fibers oriented at $0^\circ$, tested at 296~K. Fig.~\ref{fig:db_mesh} shows the quarter-symmetry geometry, the boundary conditions, and the mesh used in the phase-field simulations. The mesh comprised 1921 Q4 elements and was refined along the expected crack path. At least four elements covered the length scale $l_0$, which gave an average element size of 0.125~mm in that region. The objective function was defined as the root-sum-of-squares error between the measured and predicted stresses at every recorded strain level. To reduce the risk of the genetic algorithm converging to a local optimum, several independent runs were performed until convergence, each starting from a population of 900.

\begin{table}[H]
	\centering
	\vspace{-30pt}
	\caption{Material and model parameters used in this study.}
	\label{tab:material_parameters}
	\begin{tabular}{llll}
		\toprule
		Parameter & Symbol & Value & Unit \\
		\midrule
		\multicolumn{4}{l}{\textit{Viscoelastic (epoxy matrix, Argon flow rule)}} \\
		Equilibrium shear modulus        & $\mu_\mathrm{eq}$        & 1300                    & MPa \\
		Non-equilibrium shear modulus    & $\mu_\mathrm{neq}$       & 170                     & MPa \\
		Bulk modulus                     & $K_v$                    & 1300                    & MPa \\
		Pre-exponential strain rate      & $\dot{\varepsilon}_0$    & $1.0447\times10^{12}$   & s$^{-1}$ \\
		Activation energy                & $\Delta H$               & $1.977\times10^{-19}$   & J \\
		Athermal shear resistance        & $\tau_0$                 & 160                     & MPa \\
		Argon exponent                   & $m$                      & 0.657                   & -- \\
		\midrule
		\multicolumn{4}{l}{\textit{Viscoplastic (threshold-based flow rule)}} \\
		Hardening coefficient            & $a$                      & 0.5                     & -- \\
		Hardening exponent               & $b$                      & 1.5                     & -- \\
		Yield threshold                  & $\sigma_0$               & 20                      & MPa \\
		\midrule
		\multicolumn{4}{l}{\textit{Phase-field fracture}} \\
		Critical energy release rate     & $G_c$                    & 3.0                     & N/mm \\
		Length scale                     & $l_0$                    & 0.5                     & mm \\
		Anisotropy parameter             & $\hat{\alpha}$           & 10                      & -- \\
		\midrule
		\multicolumn{4}{l}{\textit{Fatigue}} \\
		Fatigue activation threshold     & $D_T$                    & $1\times10^{-5}$        & -- \\
		Characteristic decay scale       & $D_c$                    & $3.5\times10^{-2}$      & -- \\
		VE--VP coupling weight           & $\beta$                  & $5\times10^{-5}$        & MPa$^{-1}$ \\
		\midrule
		\multicolumn{4}{l}{\textit{Adaptive cycle-jump}} \\
		Block stationarity tolerance     & $\varepsilon_\rho$       & 0.01                    & -- \\
		Maximum cycles per block         & $N_\mathrm{block}^\mathrm{max}$ & 10               & -- \\
		Degradation increment limit      & $\Delta D_\mathrm{max}$  & $1.25\times10^{-3}$     & -- \\
		Skip safety factor               & $\alpha$                 & 0.3                     & -- \\
		Skip fraction of elapsed cycles  & $\chi$                 & 0.1                     & -- \\
		First cycle with a jump          & $N_\mathrm{jump}^\mathrm{min}$ & 40                & -- \\
		Skip size upper bound            & $N_\mathrm{skip}^\mathrm{max}$ & $1\times10^4$     & -- \\
		Clock exponent of $\kappa$       & $m_\kappa$               & 6                       & -- \\
		Clock exponent of $\bar\kappa$   & $m_{\bar\kappa}$         & 1.5                     & -- \\
		Lower bound of age ratio         & $R_\mathrm{min}$         & 0.2                     & -- \\
		\bottomrule
	\end{tabular}
\end{table}

\begin{figure}[H]
	\centering
	{\includegraphics[scale=1]{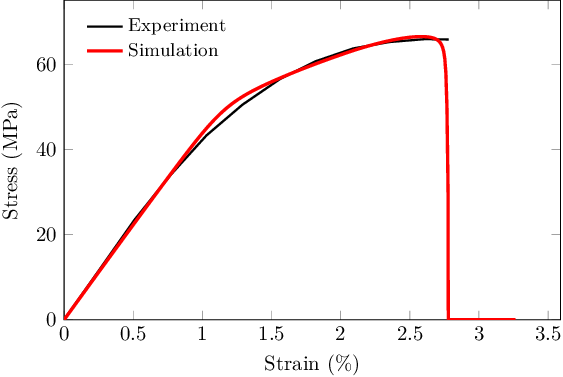}}
	\caption{Stress-strain response under monotonic tensile loading of SCFR (5 wt\% ($0^\circ$))/epoxy at 296~K. Experimental data have been taken from~\cite{hematipour2025progressive}.}
	\label{fig:dogbone_mono}
\end{figure}

Fig.~\ref{fig:dogbone_mono} compares the simulated stress-strain response with the experimental data of~\cite{hematipour2025progressive}. The model captures the initial elastic slope and the gradual loss of stiffness after the viscoplastic flow threshold $\sigma_0$ is exceeded. It also reproduces the peak stress, which is attained just below 3\% strain. Beyond this point, the load drops sharply, indicating brittle fracture as the phase-field variable reaches its critical value at the mid-section of the specimen. The good agreement holds over the entire loading history and not only at the peak load. This justifies using the calibrated VE-VP parameter set without modification in the fatigue simulations that follow.


In the second step, the VE-VP parameters were kept at their values from the first step. The fatigue parameters $\mathcal{D}_T$, $\mathcal{D}_c$, and $\beta$ were then calibrated with the experimental S-N data of~\cite{dean2026experimental} for the same material system. The cycle-jump parameters are numerical rather than material parameters and were therefore not fitted to the experiments. Instead, they were determined by a convergence analysis (Section~\ref{sec:convergence}), in which the predicted fatigue life is compared with the explicit simulation for decreasing $\Delta\mathcal{D}_\mathrm{max}$. The fatigue simulations represent tension-tension cyclic loading with a load ratio of $R_U = 0.1$. The loading is applied under displacement control as a sinusoidal waveform,
\begin{equation}
	U(t) = U_\mathrm{mean} + U_\mathrm{amp}\sin\!\left(
	\frac{2\pi t}{T_\mathrm{cyc}} - \frac{\pi}{2}\right),
	\quad
	U_\mathrm{mean} = \tfrac{1}{2}(1+R_U)U_\mathrm{max},
	\quad
	U_\mathrm{amp} = \tfrac{1}{2}(1-R_U)U_\mathrm{max},
	\label{eq:waveform}
\end{equation}
with a frequency of 1~Hz, that is, a period of $T_\mathrm{cyc}=1$~s. Owing to the phase offset, every cycle starts at the minimum displacement $U_\mathrm{min} = R_U\,U_\mathrm{max}$. For each target normalized stress level $\sigma/\sigma_\mathrm{u}$, the peak displacement $U_\mathrm{max}$ is chosen such that the calibrated monotonic response attains that stress level.

\begin{figure}[H]
	\centering
	{\includegraphics[scale=1.0]{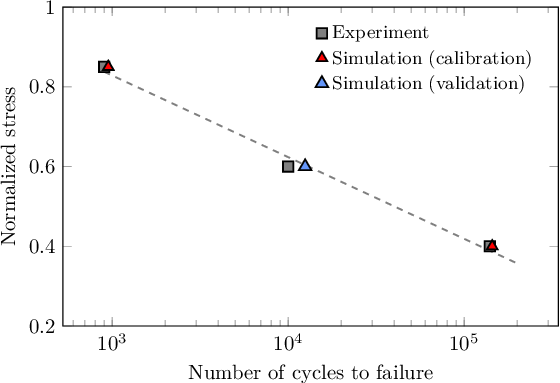}}
	\caption{S-N curve of SCFR (5 wt\% ($0^\circ$))/epoxy at 296~K. Experimental data have been taken from~\cite{dean2026experimental}. The normalized stress is $\sigma / \sigma_\text{ultimate}$.}
	\label{fig:dogbone_fatigue}
\end{figure}

Fig.~\ref{fig:dogbone_fatigue} presents the resulting S-N curve, in which the normalized stress $\sigma/\sigma_\mathrm{u}$ is plotted against the number of cycles to failure on a logarithmic axis. The simulated points at the highest and lowest stress levels, $\sigma/\sigma_\mathrm{u}=0.85$ and $0.40$, served for the calibration. The intermediate point at $\sigma/\sigma_\mathrm{u}=0.6$ was excluded from the fit and used only for validation. All three simulated points lie close to the dashed log-linear trend fitted to the experimental data. At the validation point, the model gives $N_f=12{,}488$ cycles, while the experiment shows $10{,}000$ cycles. The difference is about 0.10 decades in $\log_{10}N_f$, which is comparable to the residuals at the two calibration points. The calibrated model therefore predicts the fatigue life correctly at a stress level that was not used in the fit.


\definecolor{lightblue}{RGB}{100,150,250}

\begin{figure}[H]
	\centering
	\begin{subfigure}[b]{1\linewidth}
		\centering
		{\includegraphics[scale=1.0]{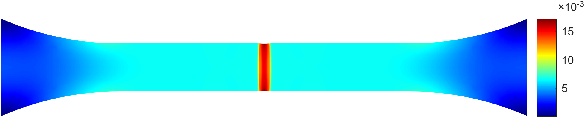}}
		\caption{ }
		\label{fig:db_kappabar1}
	\end{subfigure}
	
	\begin{subfigure}[b]{1\linewidth}
		\centering
		{\includegraphics[scale=1.0]{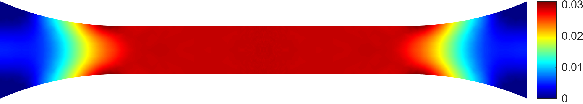}}
		\caption{ }
		\label{fig:db_kappa1}
	\end{subfigure}
	
	\begin{subfigure}[b]{1\linewidth}
		\centering
		{\includegraphics[scale=1.0]{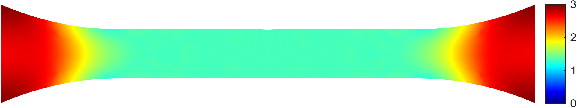}}
		\caption{ }
		\label{fig:db_Geff1}
	\end{subfigure}
	
	\begin{subfigure}[b]{1\linewidth}
		\centering
		{\includegraphics[scale=1.0]{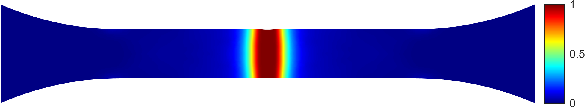}}
		\caption{ }
		\label{fig:db_phi1}
	\end{subfigure}
	
	\caption{LCF after 918 cycles at the normalized stress of 0.85: (a) Accumulated non-equilibrium strain energy ($\beta\,\bar{\kappa}$), (b) accumulated viscoplastic strain ($\kappa$), (c) effective critical energy release rate ($\mathcal{G}_c^\mathrm{eff}$), (d) phase-field damage parameter ($\phi$). Simulations were performed with $\Delta D_{max} = 0.00125$.}
	\label{fig:sen_contours_LCF}
\end{figure}

\begin{figure}[H]
	\centering
	\begin{subfigure}[b]{1\linewidth}
		\centering
		{\includegraphics[scale=1.0]{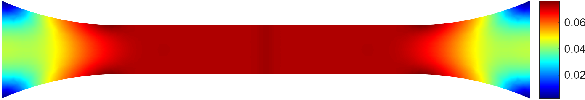}}
		\caption{ }
		\label{fig:db_kappabar2}
	\end{subfigure}
	
	\begin{subfigure}[b]{1\linewidth}
		\centering
		{\includegraphics[scale=1.0]{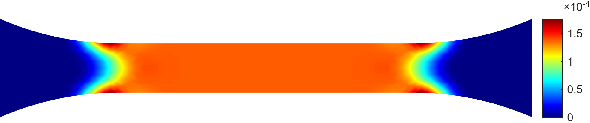}}
		\caption{ }
		\label{fig:db_kappa2}
	\end{subfigure}
	
	\begin{subfigure}[b]{1\linewidth}
		\centering
		{\includegraphics[scale=1.0]{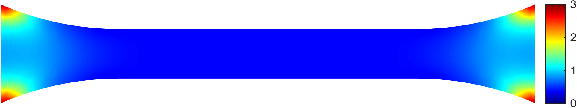}}
		\caption{ }
		\label{fig:db_Geff}
	\end{subfigure}
	
	\begin{subfigure}[b]{1\linewidth}
		\centering
		{\includegraphics[scale=1.0]{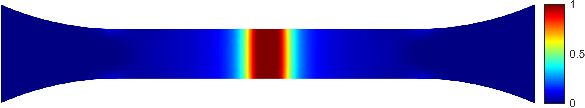}}
		\caption{ }
		\label{fig:db_phi2}
	\end{subfigure}
	
	\caption{HCF after 144014 cycles at the normalized stress of 0.4: (a) Accumulated non-equilibrium strain energy ($\beta\,\bar{\kappa}$), (b) accumulated viscoplastic strain ($\kappa$), (c) effective critical energy release rate ($\mathcal{G}_c^\mathrm{eff}$), (d) phase-field damage parameter ($\phi$). Simulations were performed with $\Delta D_{max} = 0.00125$.}
	\label{fig:sen_contours_HCF}
\end{figure}

Figs.~\ref{fig:sen_contours_LCF} and~\ref{fig:sen_contours_HCF} present the spatial distributions of four fatigue state variables at the two extreme calibration stress levels. These correspond to the low-cycle regime at $\sigma/\sigma_u=0.85$ ($N=918$ cycles) and the high-cycle regime at $\sigma/\sigma_u=0.4$ ($N=144{,}014$ cycles). The results make it possible to compare how the viscoplastic and viscoelastic mechanisms contribute to the combined driving force $\mathcal{D}_\mathrm{fat}=\kappa+\beta\bar\kappa$ (Eq.~\eqref{eq:D_fat}) in the two regimes.

In the LCF case (Fig.~\ref{fig:sen_contours_LCF}), the accumulated viscoplastic strain $\kappa$ (Fig.~\ref{fig:db_kappa1}) saturates over nearly the whole gauge section. Its values exceed those of the weighted non-equilibrium strain energy $\beta\bar\kappa$ (Fig.~\ref{fig:db_kappabar1}) by an order of magnitude, and $\beta\bar\kappa$ stays confined to a narrow band at the center of the specimen. At this high stress amplitude, the viscoplastic threshold $\tau_p>\sigma_0$ (Eq.~\eqref{eq:eps_dot_vp}) is exceeded over most of the gauge section in every cycle. As a result, $\kappa$ builds up practically wherever the material is loaded. In contrast, $\beta\bar\kappa$ grows noticeably only where the phase-field damage has already begun to localize. In this regime, the driving force $\mathcal{D}_\mathrm{fat}$, the degradation of $\mathcal{G}_c^\mathrm{eff}$ (Fig.~\ref{fig:db_Geff1}), and the resulting damage band in $\phi$ (Fig.~\ref{fig:db_phi1}) are therefore controlled almost entirely by the viscoplastic accumulator.

In the HCF case (Fig.~\ref{fig:sen_contours_HCF}), the situation is reversed. Here, the weighted non-equilibrium strain energy $\beta\bar\kappa$ (Fig.~\ref{fig:db_kappabar2}) is the field that saturates over most of the gauge section. Its values are several times larger than those of the accumulated viscoplastic strain $\kappa$ (Fig.~\ref{fig:db_kappa2}), which is both smaller and more localized. At this low stress amplitude, $\tau_p$ seldom exceeds $\sigma_0$, so the purely viscoplastic mechanism plays only a minor role. Instead, the cyclic accumulation of non-equilibrium strain energy in the sub-yield viscoelastic network, represented by $\beta\bar\kappa$, sustains $\mathcal{D}_\mathrm{fat}$. This mechanism eventually drives the degradation of $\mathcal{G}_c^\mathrm{eff}$ and the localized damage band in $\phi$ (Fig.~\ref{fig:db_phi2}). The change in the dominant mechanism between Figs.~\ref{fig:sen_contours_LCF} and~\ref{fig:sen_contours_HCF} gives a direct, spatially resolved confirmation of the model concept. The relative weight of $\kappa$ and $\beta\bar\kappa$ in $\mathcal{D}_\mathrm{fat}$ shifts naturally with the loading amplitude, even though the formulation contains no explicit distinction between fatigue regimes.


\subsection{Convergence of the cycle-jump method}
\label{sec:convergence}


Table~\ref{tab:cycle_jump_convergence} reports the effect of the jump bound $\Delta\mathcal{D}_\mathrm{max}$ on the predicted fatigue life at $\sigma/\sigma_u=0.85$ (LCF) and $0.40$ (HCF). The remaining parameters were kept at the values given in Table~\ref{tab:material_parameters}. For every bound, the table gives the number of cycles to failure $N_f$, its relative error, and the speedup. The speedup is defined as $N_f$ divided by the number of explicitly simulated cycles. For LCF, the explicit simulation of all 931 cycles serves as the reference. For HCF, an explicit simulation would require about $1.4\times10^5$ cycles and is impractical. The run with the smallest bound is therefore taken as the reference. Consequently, the HCF errors indicate the degree of convergence of the sequence rather than the deviation from an exact solution.

\begin{table}[H]
	\centering
	\caption{Cycles to failure $N_f$ for decreasing $\Delta\mathcal{D}_\mathrm{max}$. The error is relative to the explicit simulation for LCF and to the run with the smallest $\Delta\mathcal{D}_\mathrm{max}$ for HCF, for which an explicit simulation is not feasible. The speedup is $N_f$ divided by the number of explicitly simulated cycles}
	\label{tab:cycle_jump_convergence}
	\begin{tabular}{l c r r r}
		\toprule
		& $\Delta\mathcal{D}_\mathrm{max}$ & $N_f$ & Error (\%) & Speedup \\
		\midrule
		LCF & 0.01     & 1132 & $+21.6$ & 7.7 \\
		& 0.005    & 1026 & $+10.2$  & 4.8 \\
		& 0.0025   & 968  & $+4.0$   & 2.9 \\
		& 0.00125  & 918  & $-1.4$   & 1.9 \\
		& 0.000612 & 937  & $+0.6$   & 1.3 \\
		& explicit & 931  & 0       & 1 \\
		\midrule
		HCF & 0.01     & 134\,402 & $-6.7$ & 500 \\
		& 0.005    & 142\,071 & $-1.4$ & 333 \\
		& 0.0025   & 143\,404 & $-0.4$ & 200 \\
		& 0.00125  & 144\,014 & 0      & 143 \\
		\bottomrule
	\end{tabular}
\end{table}

For LCF, reducing $\Delta\mathcal{D}_\mathrm{max}$ from $0.01$ to $0.0025$ lowers the error from $+21.6\%$ to $+4.0\%$. Every halving of the bound reduces the error by a factor of about 2 to 2.5, in line with first-order convergence in $\Delta\mathcal{D}_\mathrm{max}$. Coarse bounds lead to an overestimation of the life. For $\Delta\mathcal{D}_\mathrm{max}\le0.00125$, the error remains within $1.4\%$ and changes sign, with $-1.4\%$ at $0.00125$ and $+0.6\%$ at $0.000612$. At this level, the error no longer varies systematically with the bound. The higher accuracy comes at the cost of speedup, which decreases from $7.7$ to $1.3$ as the bound is reduced. The speedup remains modest because the life is short and the damage grows rapidly. The limiter of Eq.~\eqref{eq:Nskip_correction} therefore permits only short jumps.

For HCF, the life is underestimated by $6.7\%$ at $\Delta\mathcal{D}_\mathrm{max}=0.01$, which is opposite in sign to the LCF case. The error decreases to $0.4\%$ at $0.0025$. Consecutive runs differ by $5.3\%$, $0.9\%$, and $0.4\%$, which shows that the life converges as the bound is reduced. The speedup exceeds two orders of magnitude and ranges from $500$ at the coarsest bound to $143$ at $0.00125$. In the latter run, about $10^3$ cycles are simulated explicitly instead of $1.44\times10^5$. Since the smallest bound serves as the reference, the true error of this run cannot be determined exactly. The $0.4\%$ difference from the preceding run is the best available estimate.

All simulations in this work use the bound $\Delta\mathcal{D}_\mathrm{max}=0.00125$. With this value, the error stays within $1.4\%$ in LCF, and the speedup is $1.9$ in LCF and $143$ in HCF. The method is thus most efficient in the high-cycle regime, where an explicit simulation is not feasible. In the low-cycle regime, where acceleration is not needed, it reduces almost to an explicit simulation.

\subsection{Sensitivity analysis}
\label{sec:sensitivity_analysis}

The parameters $\mathcal{D}_T$, $\mathcal{D}_c$, and $\beta$ together determine how the fatigue driving force $\mathcal{D}_\mathrm{fat}=\kappa+\beta\bar\kappa$ (Eq.~\eqref{eq:D_fat}) lowers the fracture toughness through $\mathcal{F}(\mathcal{D}_\mathrm{fat})$ (Eq.~\eqref{eq:F_deg}). Their effects on the fatigue life may thus be coupled. A one-at-a-time sweep cannot distinguish the effect of a single parameter from its interactions with the other two. With only three parameters, a full factorial study remains affordable and yields an exact decomposition of the variance of the fatigue life. The cycle-jump parameters are numerical and are examined separately in Section~\ref{sec:convergence}. Here, they were kept at their converged values.

Each parameter is assigned three levels on a logarithmic scale around its calibrated value (Table~\ref{tab:material_parameters}). The threshold $\mathcal{D}_T$ is set to $10^{-6}$, $10^{-5}$, and $10^{-4}$. The decay scale $\mathcal{D}_c$ is set to one third, one, and three times its calibrated value, namely $1.17\times10^{-2}$, $3.5\times10^{-2}$, and $1.05\times10^{-1}$. The weight $\beta$ is set to $1.67\times10^{-5}$, $5\times10^{-5}$, and $1.5\times10^{-4}$. All $3^3=27$ combinations are simulated at the two normalized stress levels $\sigma/\sigma_u=0.85$ and $0.40$. These levels represent the LCF and HCF regimes discussed in Section~\ref{sec:valid}. The response variable is $y=\ln N_f$, in line with the logarithmic axis of the S-N curve in Fig.~\ref{fig:dogbone_fatigue}.

Since each combination was simulated, the variance $V$ of $y$ over the 27 runs can be split exactly into contributions from the individual parameters and their interactions~\cite{sobol2001global},
\begin{equation}
	V = V_T + V_c + V_\beta + V_{Tc} + V_{T\beta} + V_{c\beta} + V_{Tc\beta}.
	\label{eq:V_decomp}
\end{equation}
In this expression, $V_i$ denotes the variance of the mean of $y$ over the three levels of parameter $i$. It is the variance explained by that parameter alone. The terms $V_{ij}$ and $V_{Tc\beta}$ represent the variance explained by parameter combinations in addition to their individual effects. For each parameter, two indices are reported,
\begin{equation}
	S_i = \frac{V_i}{V},
	\qquad
	S_i^\mathrm{tot} = \frac{V_i + \sum_{j\neq i} V_{ij} + V_{Tc\beta}}{V}.
	\label{eq:S_indices}
\end{equation}
The first-order index $S_i$ gives the fraction of $V$ explained by parameter $i$ alone. The total index $S_i^\mathrm{tot}$ also includes all interaction terms that involve parameter $i$. The difference $S_i^\mathrm{tot}-S_i$ is therefore the fraction contributed by parameter $i$ through interactions. The total indices may add up to more than one, since each interaction term is counted once for every parameter it contains. All three levels were assumed equally likely. The indices are thus exact for the sampled grid and free of sampling error, although they depend on the selected ranges. Because each combination was run only once, the highest-order term $V_{Tc\beta}$ also includes the numerical scatter of the cycle-jump scheme.

\begin{figure}[H]
	\centering
	{\includegraphics[scale=0.8]{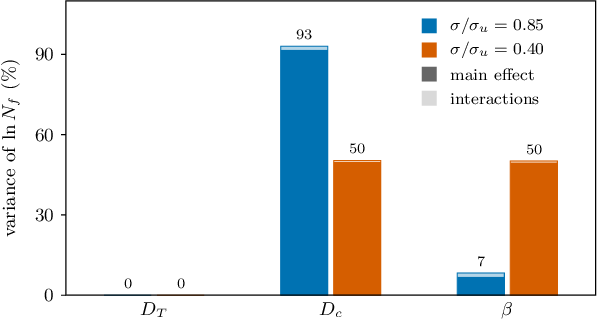}}
	\caption{Variance-based sensitivity of $\ln N_f$ to the fatigue
		parameters $D_T$, $D_c$, and $\beta$ from a $3\times3\times3$ full
		factorial study at $\sigma/\sigma_u=0.85$ (LCF) and $0.40$ (HCF). The
		full bar height is the total index $S_i^\mathrm{tot}$ and the dark part is
		the first-order index $S_i$. The light part is the share contributed
		through interactions with the other two parameters.}
	\label{fig:sensitivity_bars}
\end{figure}

Fig.~\ref{fig:sensitivity_bars} presents both indices at the two stress levels. The threshold $\mathcal{D}_T$ has no measurable influence in either regime. Across its three levels, $N_f$ varies by less than $1\%$ without a monotonic trend, which lies within the numerical scatter. This outcome is expected. The threshold enters only through $\mathcal{F}$, and its largest value is about $1\%$ of the smallest value of $\mathcal{D}_c$. It therefore changes $\mathcal{F}$ by at most about $1\%$. This conclusion is limited to the range considered, and a threshold of the same order as $\mathcal{D}_c$ would have a noticeable effect.

The decay scale $\mathcal{D}_c$ governs the fatigue life at both stress levels. At $\sigma/\sigma_u=0.85$, it explains $92\%$ of the variance alone and $93\%$ with interactions included, compared with $7\%$ and $8\%$ for $\beta$. At $\sigma/\sigma_u=0.40$, $\mathcal{D}_c$ and $\beta$ contribute almost equally, with about $50\%$ each. This shift agrees with the physical picture of Section~\ref{sec:fatigue}. At the high stress level, the viscoplastic accumulator $\kappa$ dominates $\mathcal{D}_\mathrm{fat}$, and the viscoelastic term $\beta\bar\kappa$ is small. At the low stress level, $\beta\bar\kappa$ becomes the main driver. Over the logarithmic ranges studied, $N_f$ then responds to $\mathcal{D}_c$ and $\beta$ in nearly equal and opposite proportions. The interaction shares are about $1\%$ at $\sigma/\sigma_u=0.85$ and below $1\%$ at $\sigma/\sigma_u=0.40$. The three parameters thus act on $\ln N_f$ almost independently. A change in any one of them alters the fatigue life by the same proportion, regardless of the values of the other two. This makes it possible to calibrate $\mathcal{D}_c$ and $\beta$ separately in the two regimes.

\subsection{Single-edge notched simulations}
\label{sec:sen_simulation}

%

\begin{figure}[H]
	\centering
	{\includegraphics[scale=1]{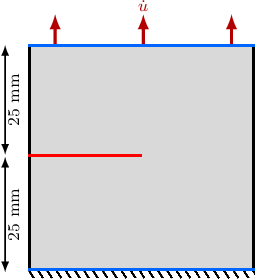}}
	\caption{Single-edge notched specimen.}
	\label{fig:sen_specimen}
\end{figure}

Fig.~\ref{fig:sen_specimen} depicts the single-edge notched specimen considered in the rest of this section. It consists of a $50\times50$~mm plate with a pre-crack that runs from the left edge to the middle of the width. The plate is loaded in tension by a displacement prescribed at the top edge, while the bottom edge is fixed. The fiber orientation differs between the configurations studied here and may deflect the crack away from the straight ligament. A mesh refined along a single expected crack path would therefore not be suitable for all cases. The mesh instead comprises 8858 Q4 elements with uniform refinement over the entire region that the crack could reach in any of the tested fiber configurations. In this region, at least four elements span the length scale $l_0$.

The fatigue loading is applied under displacement control using the sinusoidal waveform of Eq.~\eqref{eq:waveform}, with $R_U=0$ and the same frequency of 1~Hz as in Section~\ref{sec:valid}. This setting corresponds to pulsating tension. The applied displacement oscillates between zero and a fixed peak value $U_\mathrm{max}$, unlike the tension-tension loading of Section~\ref{sec:valid}, where both limits are nonzero. The peak displacement $U_\mathrm{max}$ is chosen individually for each case.

\begin{figure}[H]
	\centering
	{\includegraphics[scale=1]{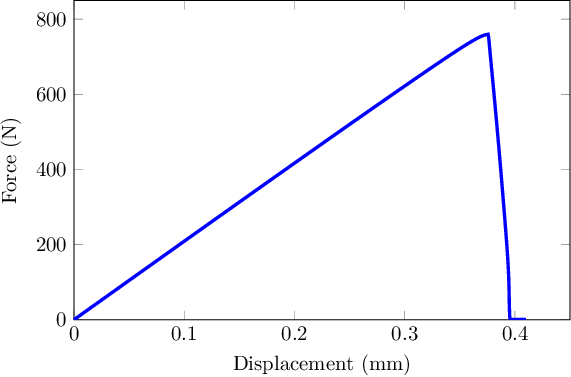}}
	\caption{Force-displacement response of the single-edge notched specimen of SCFR (5 wt\% ($0^\circ$))/epoxy at 296~K.}
	\label{fig:crack_growth_sen_05}
\end{figure}

Fig.~\ref{fig:crack_growth_sen_05} shows the monotonic
force-displacement response of the SCFR (5~wt\%, $0^\circ$) specimen
at 296~K, calibrated against the same material parameters as
Section~\ref{sec:valid}. The response is linear until crack
propagation initiates from the notch tip, after which the reaction
force drops sharply to zero once the phase-field variable reaches its
critical value along the ligament, consistent with the brittle,
unstable crack growth expected for a sharp pre-crack under
displacement control.


\begin{figure}[H]
	\centering
	\begin{subfigure}[b]{1\textwidth}
		\centering
		{\includegraphics[scale=1]{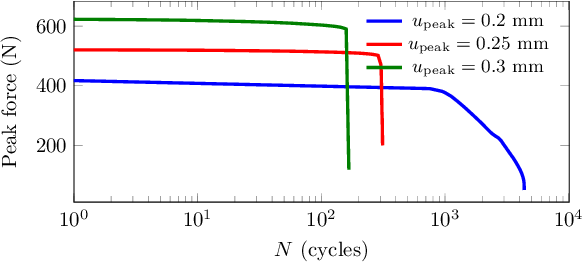}}
		\caption{}
		\label{fig:Fpeak_N_upeak}
	\end{subfigure}
	\begin{subfigure}[b]{1\textwidth}
		\centering
		{\includegraphics[scale=1]{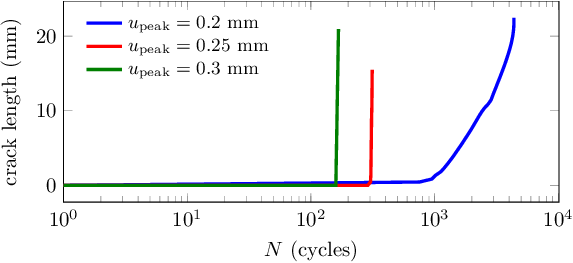}}
		\caption{}
		\label{fig:crack_upeak}
	\end{subfigure}
	\begin{subfigure}[b]{1\textwidth}
		\centering
		{\includegraphics[scale=1]{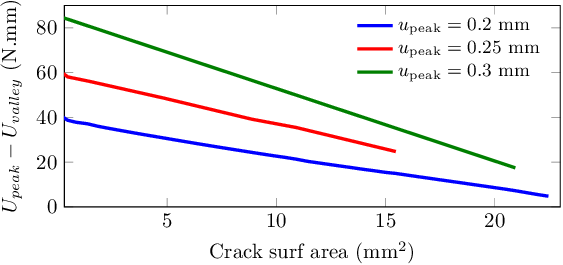}}
		\caption{}
		\label{fig:dU_upeak}
	\end{subfigure}
	\caption{Effect of the peak displacement on the responses of SCFR (5 wt\% ($0^\circ$))/epoxy under fatigue loading at 296~K: (a) Peak force vs. cycles, (b) crack length vs. cycles, and (c) dU versus crack surface area.}
	\label{fig:sen_upeak}
\end{figure}

\begin{figure}[H]
	\centering
	{\includegraphics[scale=0.8]{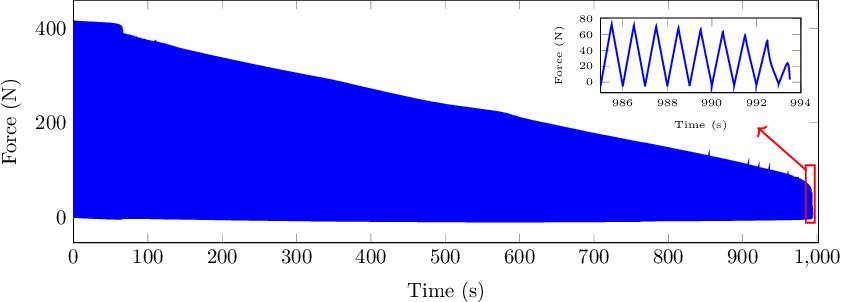}}
	\caption{Force versus time response of SCFR (5 wt\% ($0^\circ$))/epoxy at 296~K, with a zoomed inset of the $[t_1,t_2] = [380,397]$~s window (red rectangle) shown in the top-right corner. $u_\text{peak} = 0.2$ mm.}
	\label{fig:force_time}
\end{figure}

\begin{figure}[H]
	\centering
	\begin{subfigure}[b]{0.48\linewidth}
		\centering
		{\includegraphics[scale=1]{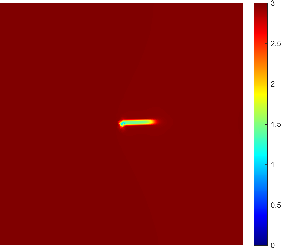}}
		\caption{ }
		\label{fig:sen_Geff1}
	\end{subfigure}
	\begin{subfigure}[b]{0.48\linewidth}
		\centering
		{\includegraphics[scale=1]{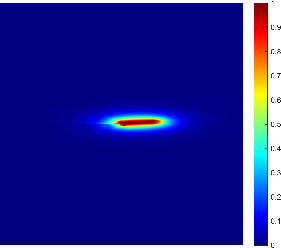}}
		\caption{ }
		\label{fig:sen_phi1}
	\end{subfigure}
	
	\begin{subfigure}[b]{0.48\linewidth}
		\centering
		{\includegraphics[scale=1]{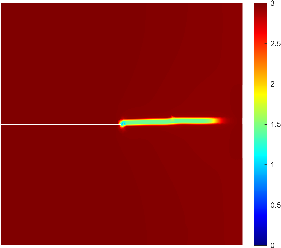}}
		\caption{ }
		\label{fig:sen_Geff2}
	\end{subfigure}
	\begin{subfigure}[b]{0.48\linewidth}
		\centering
		{\includegraphics[scale=1]{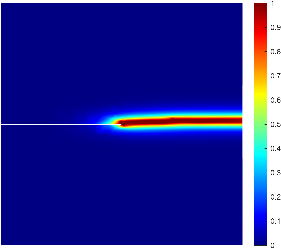}}
		\caption{ }
		\label{fig:sen_phi2}
	\end{subfigure}
	
	\caption{Fatigue fracture propagation in SCFR (5 wt\% ($0^\circ$))/epoxy at 296~K: (a) Effective critical energy release rate after 2000 cycles, (b) phase-field parameter after 2000 cycles, (c) effective critical energy release rate after 4377 cycles, (d) phase-field damage parameter after 4377 cycles. $u_\text{peak} = 0.2$ mm.}
	\label{fig:sen_contours_0degree}
\end{figure}

Fig.~\ref{fig:sen_upeak} shows the fatigue response for three peak displacements, $u_\mathrm{peak}=0.2$, $0.25$, and $0.3$~mm, at 296~K. Higher $u_\mathrm{peak}$ produces a higher initial peak force
(Fig.~\ref{fig:Fpeak_N_upeak}) but a substantially shorter fatigue life. The crack length (Fig.~\ref{fig:crack_upeak}) remains near-stationary for an incubation period before propagating rapidly to
failure, and this incubation period shortens by roughly an order of magnitude between $u_\mathrm{peak}=0.2$~mm and $u_\mathrm{peak}=0.3$~mm. Fig.~\ref{fig:dU_upeak} shows that the hysteresis dissipation $dU$ decreases monotonically with accumulated crack surface area at all three amplitudes, consistent with the crack progressively unloading the specimen as it advances. Fig.~\ref{fig:force_time} shows the force-time history at $u_\mathrm{peak}=0.2$~mm, plotting only the cycles that are explicitly simulated within each block of the adaptive cycle-jump strategy of Section~\ref{sec:cycle_jump}, rather than a continuous record of every cycle up to failure. The zoomed inset over cycles shows the cycle-by-cycle force amplitude still well resolved by the cycle-jump scheme late into the simulation, with the gradual reduction in peak force per cycle visible directly against the early-cycle response. Fig.~\ref{fig:sen_contours_0degree} shows the corresponding effective critical energy release rate and phase-field damage parameter at $N=2000$ and $N=4377$ cycles. The crack advances horizontally from the notch tip along the ligament, and the degraded $\mathcal{G}_c^\mathrm{eff}$ region ahead of the crack tip widens between the two cycle counts as the process zone develops.


\begin{figure}[H]
	\centering
	\begin{subfigure}[b]{1\textwidth}
		\centering
		{\includegraphics[scale=1]{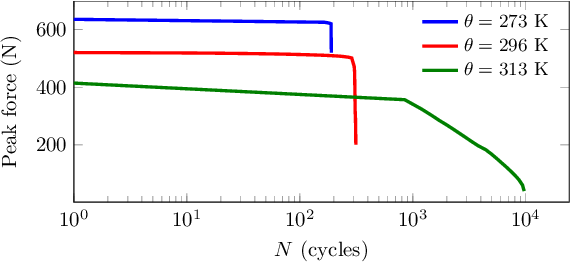}}
		\caption{}
		\label{fig:Fpeak_N_temp}
	\end{subfigure}
	\begin{subfigure}[b]{1\textwidth}
		\centering
		{\includegraphics[scale=1]{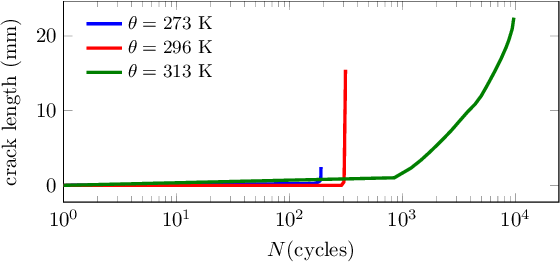}}
		\caption{}
		\label{fig:crack_temp}
	\end{subfigure}
	\begin{subfigure}[b]{1\textwidth}
		\centering
		{\includegraphics[scale=1]{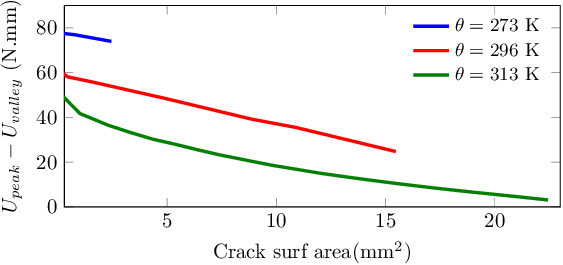}}
		\label{fig:dU_temp}
	\end{subfigure}
	\caption{Effect of temperature on the responses of SCFR (5 wt\% ($0^\circ$))/epoxy under fatigue loading: (a) Peak force vs. cycles, (b) crack length vs. cycles, and (c) dU versus crack surface area. $u_\text{peak} = 0.25$ mm.}
	\label{fig:sen_temp}
\end{figure}

Fig.~\ref{fig:sen_temp} presents the fatigue response at $\theta=273$, $296$, and $313$~K for $u_\mathrm{peak}=0.25$~mm. The lowest temperature gives the highest peak force and the shortest fatigue life. Conversely, the highest temperature gives the lowest peak force but the longest life, which exceeds that of the other two cases by an order of magnitude or more. This trend agrees with the thermally activated Argon flow rule of Eq.~\eqref{eq:eps_dot_v}. At lower temperature, the non-equilibrium network relaxes more slowly, and the material responds in a more glassy and brittle manner. At higher temperature, faster stress relaxation postpones the onset of significant crack growth.


\vspace{24pt}

\begin{figure}[H]
	\centering
	\begin{subfigure}[b]{1\textwidth}
		\centering
		{\includegraphics[scale=1]{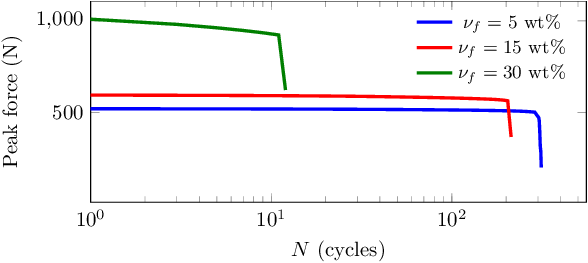}}
		\caption{}
		\label{fig:Fpeak_N_vf}
	\end{subfigure}
	\begin{subfigure}[b]{1\textwidth}
		\centering
		{\includegraphics[scale=1]{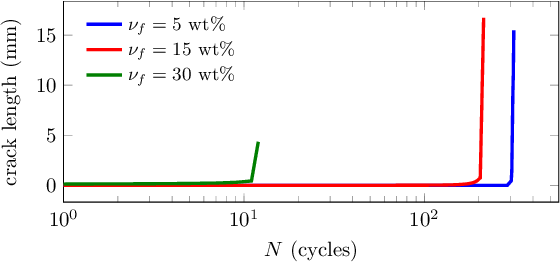}}
		\caption{}
		\label{fig:crack_vf}
	\end{subfigure}
	\begin{subfigure}[b]{1\textwidth}
		\centering
		{\includegraphics[scale=1]{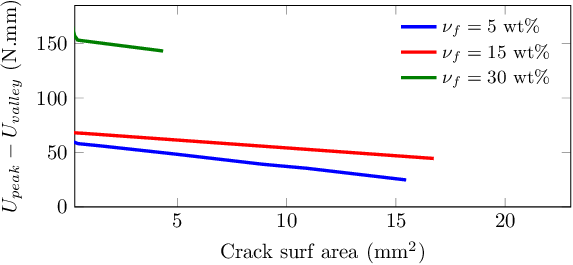}}
		\label{fig:dU_vf}
	\end{subfigure}
	\caption{Effect of fiber content on the responses of SCFR ($0^\circ$)/epoxy under fatigue loading at 296~K: (a) Peak force vs. cycles, (b) crack length vs. cycles, and (c) dU versus crack surface area. $u_\text{peak} = 0.25$ mm.}
	\label{fig:sen_vf}
\end{figure}

Fig.~\ref{fig:sen_vf} presents the fatigue response for fiber contents of $\nu_f=5$, $15$, and $30$~wt\% at $u_\mathrm{peak}=0.25$~mm and 296~K. A higher fiber content considerably increases the initial peak force, in agreement with the stiffening contribution of the fiber families in Eq.~\eqref{eq:psi_fiber}. At the same time, it strongly shortens the fatigue life. The $30$~wt\% specimen fails after about ten cycles, whereas the $5$~wt\% specimen withstands several hundred. This trade-off between monotonic stiffness and fatigue durability can be attributed to the higher local stress concentrations that a stiffer, more fiber-dominated composite carries for a given applied displacement.

\begin{figure}[H]
	\centering
	\begin{subfigure}[b]{1\textwidth}
		\centering
		{\includegraphics[scale=1]{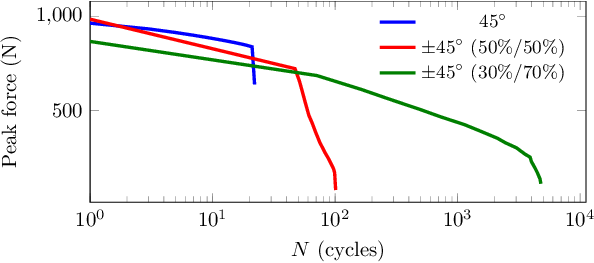}}	
		\caption{}
		\label{fig:Fpeak_N_fiberconfig}
	\end{subfigure}
	\begin{subfigure}[b]{1\textwidth}
		\centering
		{\includegraphics[scale=1]{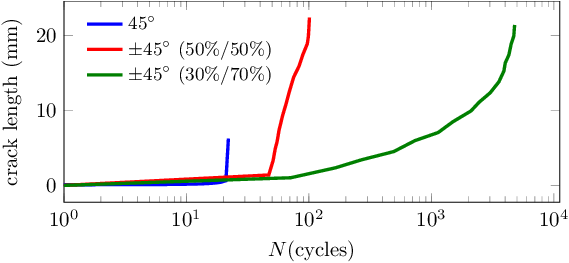}}		
		\caption{}
		\label{fig:crack_fiberconfig}
	\end{subfigure}
	\begin{subfigure}[b]{1\textwidth}
		\centering
		{\includegraphics[scale=1]{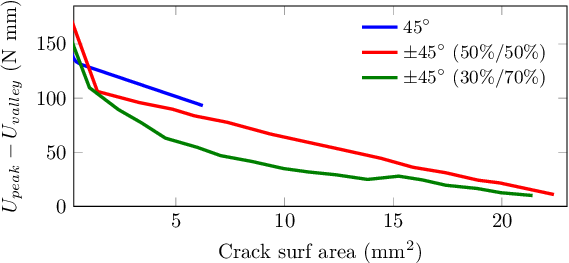}}		
		\label{fig:dU_fiberconfig}
	\end{subfigure}
	\caption{Effect of fiber orientation configuration on the responses of SCFR (30 wt\%)/epoxy under fatigue loading at 296~K: (a) Peak force vs. cycles, (b) crack length vs. cycles, and (c) dU versus crack surface area.}
	\label{fig:sen_fiberconfig}
\end{figure}

\begin{figure}[H]
	\centering
	\begin{subfigure}[b]{0.48\linewidth}
		\centering
		{\includegraphics[scale=1]{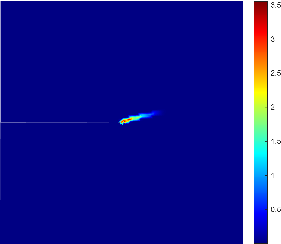}}				
		\caption{ }
		\label{fig:sen_45_betakappabar1}
	\end{subfigure}
	\begin{subfigure}[b]{0.48\linewidth}
		\centering
		{\includegraphics[scale=1]{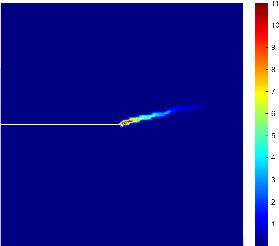}}				
		\caption{ }
		\label{fig:sen_45_betakappabar2}
	\end{subfigure}
	
	\begin{subfigure}[b]{0.48\linewidth}
		\centering
		{\includegraphics[scale=1]{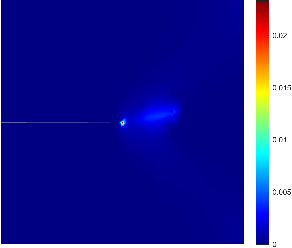}}				
		\caption{ }
		\label{fig:sen_45_kappa1}
	\end{subfigure}
	\begin{subfigure}[b]{0.48\linewidth}
		\centering
		{\includegraphics[scale=1]{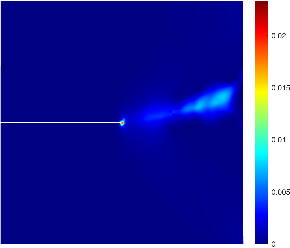}}					
		\caption{ }
		\label{fig:sen_45_kappa2}
	\end{subfigure}
	
	\begin{subfigure}[b]{0.48\linewidth}
		\centering
		{\includegraphics[scale=1]{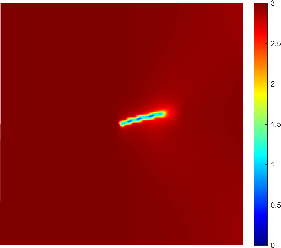}}				
		\caption{ }
		\label{fig:sen_45_Geff1}
	\end{subfigure}
	\begin{subfigure}[b]{0.48\linewidth}
		\centering
		{\includegraphics[scale=1]{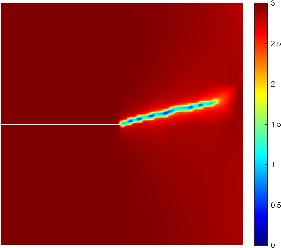}}				
		\caption{ }
		\label{fig:sen_45_Geff2}
	\end{subfigure}
	
	\begin{subfigure}[b]{0.48\linewidth}
		\centering
		{\includegraphics[scale=1]{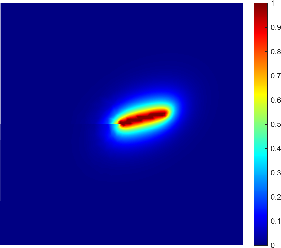}}				
		\caption{ }
		\label{fig:sen_45_phi1}
	\end{subfigure}
	\begin{subfigure}[b]{0.48\linewidth}
		\centering
		{\includegraphics[scale=1]{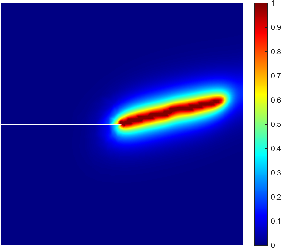}}				
		\caption{ }
		\label{fig:sen_45_phi2}
	\end{subfigure}
	
	\caption{$\pm45^\circ$ (30\%/70\%), $\nu_f=30$~wt\%: accumulated non-equilibrium strain energy $\beta\bar\kappa$ after (a) 2002 and (b) 4780 cycles; accumulated viscoplastic strain $\kappa$ after (c) 2002 and (d) 4780 cycles; effective critical energy release rate $\mathcal{G}_c^\mathrm{eff}$ after (e) 2002 and (f) 4780 cycles; phase-field damage parameter $\phi$ after (g) 2002 and (h) 4780 cycles. $u_\text{peak} = 0.38$ mm.}
	\label{fig:sen_contours}
\end{figure}

Fig.~\ref{fig:sen_fiberconfig} compares three fiber configurations at $\nu_f=30$~wt\%, $u_\mathrm{peak}=0.38$~mm, and 296~K. These are a single $45^\circ$ family, a balanced $\pm45^\circ$ (50\%/50\%) arrangement, and an unbalanced $\pm45^\circ$ (30\%/70\%) arrangement. The single $45^\circ$ and the balanced $\pm45^\circ$ configurations fail within about $100$ cycles of each other. In contrast, the unbalanced $\pm45^\circ$ (30\%/70\%) configuration lasts about two orders of magnitude longer. At this fiber content and amplitude, the asymmetric fiber split thus offers considerably higher resistance to fatigue crack growth than the other two configurations. For this longest-lasting configuration, Fig.~\ref{fig:sen_contours} shows the accumulated non-equilibrium strain energy $\beta\bar\kappa$, the accumulated viscoplastic strain $\kappa$, the effective critical energy release rate $\mathcal{G}_c^\mathrm{eff}$, and the phase-field damage parameter $\phi$ at $N=2002$ and $N=4780$ cycles.


\section{Summary and conclusions}
\label{sec:summary}


In this work, an anisotropic phase-field framework was developed for fatigue fracture of SCFRP composites at finite deformation. Within a VE-VP constitutive model, it couples the cyclic accumulation of viscoplastic strain with the accumulation of viscoelastic energy. Two fatigue history variables are introduced, namely the accumulated viscoplastic strain $\kappa$ and the accumulated non-equilibrium strain energy $\bar\kappa$. They are combined into the driving force $\mathcal{D}_\mathrm{fat}=\kappa+\beta\bar\kappa$, which reduces the fracture toughness. Since the model has no switch between regimes, the share of each mechanism in the fatigue damage is determined by the constitutive response and the load level. An adaptive cycle-jump method makes it feasible to simulate the number of cycles relevant to engineering fatigue life. The method evaluates the fatigue rates in a short explicit block, whose length follows from a stationarity criterion on the damage increment per cycle. It then extrapolates these rates over the skipped cycles and bounds the toughness drop in each jump. The main findings are summarized below.

\begin{itemize}
	\item The contributions of the two fatigue accumulators to $\mathcal{D}_\mathrm{fat}$ reverse between the loading regimes, although the model contains no explicit distinction between them. At the low-cycle stress level, the viscoplastic strain $\kappa$ saturates over almost the whole gauge section. At the high-cycle stress level, the weighted term $\beta\bar\kappa$ instead saturates the gauge section, whereas $\kappa$ stays smaller and more localized.
	
	\item The variance decomposition of $\ln N_f$ reveals that the toughness decay scale $\mathcal{D}_c$ dominates the low-cycle life and alone explains $92\%$ of the variance. The weight $\beta$ accounts for $7\%$ in this regime. At the high-cycle stress level, its share rises to $50\%$, and it controls the life jointly with $\mathcal{D}_c$. This increase is consistent with the reversal of the accumulators described above. Parameter interactions contribute about $1\%$ or less, so the parameters influence $\ln N_f$ almost independently. Within the studied range, the threshold $\mathcal{D}_T$ has no measurable effect, since its values do not exceed about $1\%$ of $\mathcal{D}_c$.
	
	\item The cycle-jump method approaches the explicit simulation as $\Delta\mathcal{D}_\mathrm{max}$ is reduced. For the value adopted in this work, $\Delta\mathcal{D}_\mathrm{max}=0.00125$, the low-cycle life deviates from the explicit result by $1.4\%$. The speedup is $1.9$ in the low-cycle regime and $143$ in the high-cycle regime, where an explicit simulation is not feasible. With a coarse bound of $0.01$, the low-cycle life is overestimated by $21.6\%$.
	
	\item The fatigue life decreases with increasing loading amplitude and increases strongly with temperature, in agreement with the thermally activated Argon flow rule. At lower temperature, the specimens carry a higher peak force but relax more slowly. They therefore respond in a more brittle manner and fail after considerably fewer cycles.
	
	\item The fatigue life drops sharply as the fiber content increases, even though the monotonic stiffness and the peak force rise. This reflects the higher local stress concentrations at a given applied displacement. The fiber configuration has an equally strong influence. At the same fiber content, an unbalanced $\pm45^\circ$ distribution withstands considerably more cycles than a single off-axis family or a balanced $\pm45^\circ$ distribution.
\end{itemize}

Several directions remain for future research. Variable-amplitude load spectra could be treated with a cycle-counting method such as rainflow counting. The fatigue formulation could be extended to hygrothermal service conditions and applied to continuous and laminated fiber architectures. Further experimental validation over a wider range of stress levels and material systems is also needed.

\section*{Acknowledgements}
The authors acknowledge the resources and support provided by the Norwegian
Research Infrastructure Services (NRIS).
The computational work was performed on resources provided through the Sigma2
national e-infrastructure, project NN10041K.

\appendix
\section{Fiber strain energy and stress}
\label{app:fiber}

The fiber contribution of family $i$ follows~\cite{arash2019viscoelastic2}
\begin{align}
	\psi_{\text{fiber}}^{(i)}
	= \frac{1}{2}\mu\bigg[&
	\!\left(v_m + v_f^{(i)}f\!\left(\bar{I}_4^{(i)}\right)\right)
	\!\left(\bar{I}_4^{(i)} + 2\left(\bar{I}_4^{(i)}\right)^{-1/2} - 3\right)
	\nonumber\\
	&+ g_1\!\left(\bar{I}_4^{(i)}\right)
	\!\left(\bar{I}_5^{(i)} - \left(\bar{I}_4^{(i)}\right)^2\right)
	\!\left(\bar{I}_4^{(i)}\right)^{-1}
	\nonumber\\
	&+ g_2\!\left(\bar{I}_4^{(i)}\right)
	\!\left(\bar{I}_1 -
	\!\left(\bar{I}_5^{(i)} + 2\left(\bar{I}_4^{(i)}\right)^{1/2}\right)
	\!\left(\bar{I}_4^{(i)}\right)^{-1}\right)\bigg],
	\label{eq:psi_fiber}
\end{align}
where $\mu$ stands for $\mu_{\text{eq}}$ or $\mu_{\text{neq}}$, and the
invariants are
\begin{equation}
	\bar{I}_1 = \operatorname{tr}[\bar{\mathbf{B}}],
	\qquad
	\bar{I}_4^{(i)} = \mathbf{a}_0^{(i)} \cdot \bar{\mathbf{C}}\,\mathbf{a}_0^{(i)},
	\qquad
	\bar{I}_5^{(i)} = \mathbf{a}_0^{(i)} \cdot \bar{\mathbf{C}}^2\mathbf{a}_0^{(i)}.
	\label{eq:I4}
\end{equation}
Here $\bar{\mathbf{C}}$ and $\bar{\mathbf{B}}$ are the tensors of the branch
under consideration, that is, $\bar{\mathbf{C}}^{\text{ve}}$ and
$\bar{\mathbf{B}}^{\text{ve}}$ in $\psi_{\text{eq}}$, and $\bar{\mathbf{C}}^e$
and $\bar{\mathbf{B}}^e$ in $\psi_{\text{neq}}$. The invariant
$\bar{I}_4^{(i)}$ is the squared fiber stretch. The stiffening functions are
\begin{equation}
	f(\bar{I}_4) = a_1 + a_2\exp\!\left[a_3(\bar{I}_4 - 1)\right],
	\label{eq:f_function}
\end{equation}
\begin{equation}
	g_1(\bar{I}_4) =
	\frac{(1 + v_f^{(i)})f(\bar{I}_4) + (1 - v_f^{(i)})}
	{(1 - v_f^{(i)})f(\bar{I}_4) + 1 + v_f^{(i)}},
	\qquad
	g_2(\bar{I}_4) =
	\frac{(1 + 0.4v_f^{(i)})f(\bar{I}_4) + 0.4(1 - v_f^{(i)})}
	{(1 - v_f^{(i)})f(\bar{I}_4) + 0.4 + v_f^{(i)}},
	\label{eq:g1_function}
\end{equation}
where $a_1$, $a_2$, and $a_3$ are material constants calibrated from uniaxial
tension data.

The fiber stress of family $i$ is
\begin{align}
	\boldsymbol{\sigma}_{\text{fiber}}^{(i)}
	= \frac{2}{J}\bigg[&
	W_1^{(i)}\operatorname{dev}[\bar{\mathbf{B}}]
	+ W_4^{(i)}\bar{I}_4^{(i)}
	\!\left(\mathbf{a} \otimes \mathbf{a} - \tfrac{1}{3}\mathbf{I}\right)
	\nonumber\\
	&+ W_5^{(i)}\bar{I}_4^{(i)}
	\!\left(\mathbf{a}\otimes\bar{\mathbf{B}}\mathbf{a}
	+ \bar{\mathbf{B}}\mathbf{a}\otimes\mathbf{a}
	- \tfrac{2}{3}\bar{I}_5^{(i)}\mathbf{I}\right)\bigg],
	\label{eq:sigma_fiber}
\end{align}
where $\mathbf{a} = \bar{\mathbf{F}}\mathbf{a}_0^{(i)}/(J^{2/3}\bar{I}_4^{(i)})^{1/2}$
is the unit fiber direction in the current configuration, and
\begin{equation}
	W_1^{(i)} = \tfrac{1}{2}\mu\, g_2,
	\qquad
	W_5^{(i)} = \frac{\mu}{2\bar{I}_4}(g_1 - g_2),
	\label{eq:W1}
\end{equation}
\begin{align}
	W_4^{(i)} = \frac{1}{2}\mu\bigg[&
	v_f^{(i)}f'\!\left(\bar{I}_4 + 2\bar{I}_4^{-1/2} - 3\right)
	+ (v_m + v_f^{(i)}f)(1 - \bar{I}_4^{-3/2})
	\nonumber\\
	&- g_1(\bar{I}_5\bar{I}_4^{-2} + 1)
	+ g_2(\bar{I}_5\bar{I}_4^{-2} + \bar{I}_4^{-3/2})
	\nonumber\\
	&+ \frac{\bar{I}_5 - \bar{I}_4^2}{2\bar{I}_4}g_1'
	+ \frac{1}{2}\!\left(\bar{I}_1
	- \frac{\bar{I}_5 + 2\bar{I}_4^{1/2}}{\bar{I}_4}\right)g_2'\bigg].
	\label{eq:W4}
\end{align}
Primes denote derivatives with respect to $\bar{I}_4^{(i)}$.

\section{Time integration of the internal variables}
\label{app:integration}

The viscous and viscoplastic velocity gradients are obtained by pulling the
stretching tensors of Eqs.~\eqref{eq:eps_dot_v} and~\eqref{eq:eps_dot_vp} back to
the intermediate configurations,
\begin{equation}
	\tilde{\mathbf{L}}^v = (\bar{\mathbf{F}}^e)^{-1}\mathbf{D}^v\bar{\mathbf{F}}^e,
	\qquad
	\tilde{\mathbf{L}}^{\text{vp}} = (\bar{\mathbf{F}}^{\text{ve}})^{-1}\mathbf{D}^{\text{vp}}\bar{\mathbf{F}}^{\text{ve}},
	\label{eq:Lve_decomp}
\end{equation}
where the viscous and viscoplastic spins are set to zero so that the
intermediate configurations are unique~\cite{qi2005stress}. The internal
variables then evolve as $\dot{\bar{\mathbf{F}}}^v=\tilde{\mathbf{L}}^v\bar{\mathbf{F}}^v$
and $\dot{\bar{\mathbf{F}}}^{\text{vp}}=\tilde{\mathbf{L}}^{\text{vp}}\bar{\mathbf{F}}^{\text{vp}}$, or explicitly
\begin{equation}
	\dot{\bar{\mathbf{F}}}^v
	= (\bar{\mathbf{F}}^e)^{-1}
	\frac{\dot{\varepsilon}^v}{\tau_{\text{neq}}}
	\operatorname{dev}\!\left[\boldsymbol{\sigma}'_{\text{neq}}\right]
	\bar{\mathbf{F}}^{\text{ve}},
	\qquad
	\dot{\bar{\mathbf{F}}}^{\text{vp}}
	= (\bar{\mathbf{F}}^{\text{ve}})^{-1}
	\frac{\dot{\varepsilon}^{\text{vp}}}{\tau_{\text{p}}}
	\operatorname{dev}[\boldsymbol{\sigma}]\,\bar{\mathbf{F}}.
	\label{eq:Fv_evol}
\end{equation}
These are first-order ODEs whose right-hand sides depend on the stress at
$t_{n+1}$, which itself depends on the unknown internal variables. A forward
Euler update would not preserve $\det(\bar{\mathbf{F}}^v)=\det(\bar{\mathbf{F}}^{\text{vp}})=1$.
An exponential map integrator~\cite{simo1992algorithms} is therefore used. For
either internal gradient,
\begin{equation}
	\bar{\mathbf{F}}_{n+1}
	= \exp\!\bigl(\Delta t\,\tilde{\mathbf{L}}_{n+1}\bigr)\,\bar{\mathbf{F}}_n.
	\label{eq:exp_map}
\end{equation}
Since the flow rules give $\operatorname{tr}(\tilde{\mathbf{L}}^v)=\operatorname{tr}(\tilde{\mathbf{L}}^{\text{vp}})=0$,
$\det(\exp(\Delta t\,\tilde{\mathbf{L}}))=\exp(\Delta t\operatorname{tr}\tilde{\mathbf{L}})=1$, so
the isochoric constraint holds exactly at every iteration. Because
$\tilde{\mathbf{L}}_{n+1}$ depends on the stress at $t_{n+1}$, the update is
solved by the fixed-point iteration of Algorithm~\ref{alg:time_integration}.

\begin{algorithm}[H]
	\caption{Implicit time integration over $[t_n,t_{n+1}]$ with fixed-point
		iteration.}
	\label{alg:time_integration}
	\begin{algorithmic}[1]
		\State \textbf{Given} $\mathbf{F}_{n+1}$, $\bar{\mathbf{F}}^v_n$,
		$\bar{\mathbf{F}}^{\text{vp}}_n$, time step $\Delta t$, tolerance $\varepsilon_{\text{tol}}$.
		\State Initialize $\bar{\mathbf{F}}^v = \bar{\mathbf{F}}^v_n$,
		$\bar{\mathbf{F}}^{\text{vp}} = \bar{\mathbf{F}}^{\text{vp}}_n$,
		\texttt{converged} $= \text{false}$.
		\While{\texttt{converged} $= \text{false}$}
		\State $J = \det(\mathbf{F}_{n+1})$,\;
		$\bar{\mathbf{F}} = J^{-1/3}\mathbf{F}_{n+1}$,\;
		$\bar{\mathbf{F}}^{\text{ve}} = \bar{\mathbf{F}}\,
		(\bar{\mathbf{F}}^{\text{vp}})^{-1}$,\;
		$\bar{\mathbf{F}}^{e} = \bar{\mathbf{F}}^{\text{ve}}\,
		(\bar{\mathbf{F}}^v)^{-1}$.
		\State Compute the Cauchy stress $\boldsymbol{\sigma}$ from
		Eqs.~\eqref{eq:sigma_total}--\eqref{eq:sigma_matrix} and~\ref{app:fiber}.
		\State Polar decomposition $\bar{\mathbf{F}}^{e} = \mathbf{R}_e\mathbf{U}_e$ and
		$\boldsymbol{\sigma}_{\mathrm{neq}}' =
		\mathbf{R}_e^T\,\boldsymbol{\sigma}_{\mathrm{neq}}\,\mathbf{R}_e$.
		\State Viscous update. Compute $\tau_{\mathrm{neq}}$,
		$\dot{\varepsilon}^v$ from Eq.~\eqref{eq:eps_dot_v},
		$\mathbf{D}^v$, $\tilde{\mathbf{L}}^v$, and
		$\bar{\mathbf{F}}^v_{\text{new}} =
		\exp\!\bigl(\Delta t\,\tilde{\mathbf{L}}^v\bigr)\bar{\mathbf{F}}^v$.
		\State Viscoplastic update. Compute $\tau_{\mathrm{p}}$.
		\If{$\tau_{\mathrm{p}} \geq \sigma_0$}
		\State Compute $\dot{\varepsilon}^{\text{vp}}$ from
		Eq.~\eqref{eq:eps_dot_vp}, $\mathbf{D}^{\text{vp}}$, $\tilde{\mathbf{L}}^{\text{vp}}$, and
		$\bar{\mathbf{F}}^{\text{vp}}_{\text{new}} =
		\exp\!\bigl(\Delta t\,\tilde{\mathbf{L}}^{\text{vp}}\bigr)
		\bar{\mathbf{F}}^{\text{vp}}$.
		\Else
		\State $\bar{\mathbf{F}}^{\text{vp}}_{\text{new}} =
		\bar{\mathbf{F}}^{\text{vp}}$.
		\EndIf
		\State $\Delta^v = \|\bar{\mathbf{F}}^v_{\text{new}} -
		\bar{\mathbf{F}}^v\|_F$,\;
		$\Delta^{\text{vp}} = \|\bar{\mathbf{F}}^{\text{vp}}_{\text{new}} -
		\bar{\mathbf{F}}^{\text{vp}}\|_F$.
		\State Update $\bar{\mathbf{F}}^v \leftarrow \bar{\mathbf{F}}^v_{\text{new}}$ and
		$\bar{\mathbf{F}}^{\text{vp}} \leftarrow \bar{\mathbf{F}}^{\text{vp}}_{\text{new}}$.
		\If{$\Delta^v + \Delta^{\text{vp}} < \varepsilon_{\text{tol}}$}
		\State Set $\bar{\mathbf{F}}^v_{n+1}=\bar{\mathbf{F}}^v$,\;
		$\bar{\mathbf{F}}^{\text{vp}}_{n+1}=\bar{\mathbf{F}}^{\text{vp}}$,\;
		\texttt{converged} $= \text{true}$.
		\EndIf
		\EndWhile
	\end{algorithmic}
\end{algorithm}

\section{Linearization and consistent tangent}
\label{app:fem}

The weak forms of Eqs.~\eqref{eq:weak_u_spat} and~\eqref{eq:weak_phi_spat} give
the residuals $\mathbf{r}^u=\mathbf{f}^u_{\text{int}}-\mathbf{f}^u_{\text{ext}}$
and $\mathbf{r}^\phi=\mathbf{f}^\phi_{\text{int}}$ with
\begin{align}
	\mathbf{f}^u_{\text{int}}
	&= \int_{\Omega_t} \boldsymbol{\sigma} : \nabla_x \boldsymbol{\eta}_u \, dv,
	\qquad
	\mathbf{f}^u_{\text{ext}}
	= \int_{\Omega_t} \rho_t \mathbf{b} \cdot \boldsymbol{\eta}_u \, dv
	+ \int_{\Gamma_t} \bar{\mathbf{t}} \cdot \boldsymbol{\eta}_u \, da,
	\label{eq:fint_u}\\
	\mathbf{f}^\phi_{\text{int}}
	&= \int_{\Omega_t} \!\left[
	J^{-1}g'(\phi)\mathcal{H}\eta_\phi
	+ J^{-1}\frac{\mathcal{G}_c^\mathrm{eff}}{l_0}\phi\eta_\phi
	+ J^{-1}\mathcal{G}_c^\mathrm{eff} l_0
	\nabla_x \phi \cdot \hat{\mathbf{A}} \cdot \nabla_x \eta_\phi
	\right]\! dv.
	\label{eq:fint_phi}
\end{align}
In the staggered scheme, the history field $\mathcal{H}$ is fixed in each
phase-field solve and the displacement and phase-field increments follow from
$\mathbf{K}^{uu}\Delta\mathbf{u}=-\mathbf{r}^u$ at fixed $\phi$ and
$\mathbf{K}^{\phi\phi}\Delta\boldsymbol{\phi}=-\mathbf{r}^\phi$ at fixed
$\mathbf{u}$. The off-diagonal blocks of the monolithic tangent are therefore
not needed. The two diagonal blocks are
\begin{align}
	\mathbf{K}^{uu}
	&= \int_{\Omega} \mathbf{B}_u^T \hat{\mathbf{c}}\, \mathbf{B}_u \, d\Omega
	+ \int_{\Omega} \mathbf{B}_u^T \boldsymbol{\sigma}\, \mathbf{B}_u \, d\Omega,
	\label{eq:Kuu}\\
	\mathbf{K}^{\phi\phi}
	&= \int_{\Omega} J^{-1}\!\left[
	\mathbf{N}_\phi^T \!\left( g''(\phi) \mathcal{H}
	+ \frac{\mathcal{G}_c^\mathrm{eff}}{l_0} \right)
	\mathbf{N}_\phi
	+ \mathcal{G}_c^\mathrm{eff} l_0\,
	\mathbf{B}_\phi^T \hat{\mathbf{A}}\, \mathbf{B}_\phi
	\right]\! d\Omega,
	\label{eq:Kphiphi}
\end{align}
where $\mathbf{N}_\phi$ is the shape function matrix of the phase field,
$\mathbf{B}_u$ and $\mathbf{B}_\phi$ are the gradient operators of the
displacement and the phase field, and $\hat{\mathbf{c}}$ is the spatial tangent
modulus.

A closed-form tangent is impractical for the coupled VE-VP-damage model, so the
Jaumann tangent $\mathbf{C}^{\sigma J}$ is computed numerically. The
Jaumann-Zaremba rate $\overset{\nabla}{\boldsymbol{\sigma}}=\dot{\boldsymbol{\sigma}}-\mathbf{W}\boldsymbol{\sigma}-\boldsymbol{\sigma}\mathbf{W}^T=\mathbf{C}^{\sigma J}:\mathbf{D}$
is objective under rigid body motion, where $\mathbf{D}$ and $\mathbf{W}$ are the
rate of deformation and the spin. Following~\cite{sun2008numerical}, the
components $(k,l)$ of the deformation gradient are perturbed symmetrically by
$\Delta\mathbf{F}_{kl}=\tfrac{\epsilon}{2}(\mathbf{e}_k\otimes\mathbf{e}_l\,\mathbf{F}+\mathbf{e}_l\otimes\mathbf{e}_k\,\mathbf{F})$
with $\epsilon=10^{-5}$. This perturbation has $\Delta\mathbf{W}_{kl}=\mathbf{0}$
and $\Delta\mathbf{D}_{kl}=\tfrac{\epsilon}{2}(\mathbf{e}_k\otimes\mathbf{e}_l+\mathbf{e}_l\otimes\mathbf{e}_k)$,
so the Jaumann rate reduces to $\Delta\boldsymbol{\sigma}=\mathbf{C}^{\sigma J}:\Delta\mathbf{D}$ and
\begin{equation}
	\mathbf{C}^{\sigma J}_{kl}
	= \frac{1}{\epsilon}
	\Bigl[\boldsymbol{\sigma}\!\left(\mathbf{F}+\Delta\mathbf{F}_{kl}\right)
	- \boldsymbol{\sigma}(\mathbf{F})\Bigr].
	\label{eq:CsigJ_numerical}
\end{equation}
The procedure is summarized in Algorithm~\ref{alg:tangent_algorithm}. The
spatial tangent follows from the Jaumann-to-Truesdell
correction~\cite{arash2025phase},
\begin{equation}
	\hat{\mathbf{c}} = \mathbf{C}^{\sigma J}
	- \tfrac{1}{2}\bigl(
	\mathbf{I}\,\overline{\otimes}\,\boldsymbol{\sigma}
	+ \mathbf{I}\,\underline{\otimes}\,\boldsymbol{\sigma}
	+ \boldsymbol{\sigma}\,\overline{\otimes}\,\mathbf{I}
	+ \boldsymbol{\sigma}\,\underline{\otimes}\,\mathbf{I}
	\bigr)
	+ \boldsymbol{\sigma} \otimes \mathbf{I},
	\label{eq:chat_from_CsigJ}
\end{equation}
and is symmetrized to remove floating-point noise. This numerical
differentiation avoids closed-form tangent expressions for the coupled model
and retains the quadratic convergence of the Newton-Raphson iterations.

\begin{algorithm}[H]
	\caption{Computation of the consistent spatial tangent modulus.}
	\label{alg:tangent_algorithm}
	\begin{algorithmic}[1]
		\State Set $\epsilon = 10^{-5}$ and evaluate $\boldsymbol{\sigma}(\mathbf{F})$
		with the model of Section~\ref{sec:const}.
		\For{$k = 1, \ldots, n_{\text{dim}}$}
		\For{$l = 1, \ldots, n_{\text{dim}}$}
		\State Set $\hat{\mathbf{F}} = \mathbf{F}$, $\hat{F}_{kl} \mathrel{+}= \epsilon/2$,
		$\hat{F}_{lk} \mathrel{+}= \epsilon/2$.
		\State Evaluate $\boldsymbol{\sigma}(\hat{\mathbf{F}})$ and store the columns
		$\mathbf{C}^{\sigma J}_{kl}$ and $\mathbf{C}^{\sigma J}_{lk}$ from
		Eq.~\eqref{eq:CsigJ_numerical}.
		\EndFor
		\EndFor
		\State Store $\mathbf{C}^{\sigma J}$ in Voigt notation and compute $\hat{\mathbf{c}}$
		from Eq.~\eqref{eq:chat_from_CsigJ}.
	\end{algorithmic}
\end{algorithm}

\section{Toughness-drop limit of the cycle jump}
\label{app:limiter}

This appendix derives the limit $N_\mathrm{skip}^\mathrm{lim}$ of Eq.~\eqref{eq:Nskip_lim}. Consider a point of $\Omega_{\mathcal{F}}$, that is, a point with $\dot{\mathcal{D}}_\mathrm{fat}>0$ and $\mathcal{F}>\Delta\mathcal{D}_\mathrm{max}$. The toughness fraction is
\begin{equation}
	\mathcal{F}(\mathcal{D}_\mathrm{fat})=
	\begin{cases}
		1, & \mathcal{D}_\mathrm{fat}\le\mathcal{D}_T,\\[2pt]
		\exp\!\left(-\dfrac{\mathcal{D}_\mathrm{fat}-\mathcal{D}_T}{\mathcal{D}_c}\right), & \mathcal{D}_\mathrm{fat}>\mathcal{D}_T.
	\end{cases}
	\label{eq:F_app}
\end{equation}
After $N$ skipped cycles of the linear extrapolation, the driving force and the toughness fraction are
\begin{equation}
	\mathcal{D}_N=\mathcal{D}_\mathrm{fat}+N\dot{\mathcal{D}}_\mathrm{fat},
	\qquad
	\mathcal{F}_N=\mathcal{F}(\mathcal{D}_N).
	\label{eq:FN_def}
\end{equation}
Since $\dot{\mathcal{D}}_\mathrm{fat}>0$, $\mathcal{D}_N$ increases with $N$. Since $\mathcal{F}$ does not increase with its argument, $\mathcal{F}_N$ does not increase with $N$.

Define the target value $\mathcal{F}_*=\mathcal{F}-\Delta\mathcal{D}_\mathrm{max}$, which lies in $(0,1)$ because $\mathcal{F}\le1$ and $\mathcal{F}>\Delta\mathcal{D}_\mathrm{max}$. The toughness drop stays within the bound if and only if the toughness fraction stays above this target,
\begin{equation}
	\mathcal{F}-\mathcal{F}_N\le\Delta\mathcal{D}_\mathrm{max}
	\iff
	\mathcal{F}_N\ge\mathcal{F}_* .
	\label{eq:drop_cond}
\end{equation}
Because $\mathcal{F}_*<1$, the value $\mathcal{F}_*$ is reached above the threshold. Solving $\mathcal{F}(\mathcal{D}_*)=\mathcal{F}_*$ with Eq.~\eqref{eq:F_app} gives
\begin{equation}
	\exp\!\left(-\frac{\mathcal{D}_*-\mathcal{D}_T}{\mathcal{D}_c}\right)=\mathcal{F}_*
	\quad\Longrightarrow\quad
	\mathcal{D}_*=\mathcal{D}_T-\mathcal{D}_c\ln\mathcal{F}_* .
\end{equation}
Since $\mathcal{F}$ does not increase, $\mathcal{F}(\mathcal{D}_N)\ge\mathcal{F}_*$ holds if and only if $\mathcal{D}_N\le\mathcal{D}_*$. With Eq.~\eqref{eq:FN_def} this reads
\begin{equation}
	\mathcal{D}_\mathrm{fat}+N\dot{\mathcal{D}}_\mathrm{fat}\le\mathcal{D}_T-\mathcal{D}_c\ln\!\left(\mathcal{F}-\Delta\mathcal{D}_\mathrm{max}\right).
\end{equation}
Dividing by $\dot{\mathcal{D}}_\mathrm{fat}>0$ gives the limit
\begin{equation}
	N\le N_\mathrm{skip}^\mathrm{lim}
	=\frac{\mathcal{D}_T-\mathcal{D}_\mathrm{fat}-\mathcal{D}_c\ln\!\left(\mathcal{F}-\Delta\mathcal{D}_\mathrm{max}\right)}{\dot{\mathcal{D}}_\mathrm{fat}},
	\label{eq:Nlim_derived}
\end{equation}
which is Eq.~\eqref{eq:Nskip_lim}. Together with Eq.~\eqref{eq:drop_cond} it proves
\begin{equation}
	\mathcal{F}-\mathcal{F}_N\le\Delta\mathcal{D}_\mathrm{max}
	\iff
	N\le N_\mathrm{skip}^\mathrm{lim}(\mathbf{x}).
	\label{eq:Nlim_iff}
\end{equation}
The derivation did not use $\mathcal{D}_\mathrm{fat}\ge\mathcal{D}_T$, so Eq.~\eqref{eq:Nlim_iff} also holds at points below the threshold, where $\mathcal{F}=1$.

At a point with $\mathcal{D}_\mathrm{fat}\ge\mathcal{D}_T$, Eq.~\eqref{eq:F_app} gives $\ln\mathcal{F}=-(\mathcal{D}_\mathrm{fat}-\mathcal{D}_T)/\mathcal{D}_c$, that is, $\mathcal{D}_T-\mathcal{D}_\mathrm{fat}=\mathcal{D}_c\ln\mathcal{F}$. Inserting this into Eq.~\eqref{eq:Nlim_derived} yields
\begin{equation}
	N_\mathrm{skip}^\mathrm{lim}
	=\frac{\mathcal{D}_c\ln\mathcal{F}-\mathcal{D}_c\ln(\mathcal{F}-\Delta\mathcal{D}_\mathrm{max})}{\dot{\mathcal{D}}_\mathrm{fat}}
	=-\frac{\mathcal{D}_c}{\dot{\mathcal{D}}_\mathrm{fat}}\ln\!\left(1-\frac{\Delta\mathcal{D}_\mathrm{max}}{\mathcal{F}}\right).
	\label{eq:Nlim_closed}
\end{equation}
For $\Delta\mathcal{D}_\mathrm{max}/\mathcal{F}\ll1$ we have $\ln(1-x)\approx-x$, so
\begin{equation}
	N_\mathrm{skip}^\mathrm{lim}\approx\frac{\mathcal{D}_c}{\dot{\mathcal{D}}_\mathrm{fat}}\,\frac{\Delta\mathcal{D}_\mathrm{max}}{\mathcal{F}} .
	\label{eq:Nlim_approx}
\end{equation}
The allowed jump therefore grows with $\mathcal{D}_c$ and $\Delta\mathcal{D}_\mathrm{max}$ and shrinks with the damage rate $\dot{\mathcal{D}}_\mathrm{fat}$ and with the loss of toughness.

\bibliography{mybibfile}

\end{document}